\documentclass[useAMS,usenatbib]{mn2e}
\usepackage{myaasmacros}
\usepackage{graphicx}
\usepackage[normalem]{ulem}
\usepackage{amsmath}
\usepackage{multirow}
\usepackage{color}
\usepackage{amssymb}
\usepackage{lscape}

\def\Glass{{\sc Glass}}
\def\PixeLens{{\sc PixeLens}}

\title[Light versus dark in strong-lens galaxies]{Light versus dark in strong-lens galaxies: Dark matter haloes that are rounder than their stars}

\author[Bruderer et al.]
{\parbox{\textwidth}{Claudio Bruderer,$^{1}$\thanks{E-mail: \texttt{claudio.bruderer@phys.ethz.ch}}
Justin I. Read,$^{1,2}$
Jonathan P. Coles,$^{3,4}$
Dominik Leier,$^{5}$
Emilio E. Falco,$^{6}$
Ignacio Ferreras$^{7}$ and
Prasenjit Saha$^{4,8}$}\vspace{0.4cm}\\
\parbox{\textwidth}{$^{1}$Institute for Astronomy, Department of Physics, ETH Zurich, Wolfgang-Pauli-Strasse 27, 8093 Z\"urich, Switzerland\\
$^{2}$Department of Physics, University of Surrey, Guildford, GU2 7XH, UK\\
$^{3}$Exascale Research Computing Lab, Campus Teratec, 2 Rue de la Piquetterie, 91680 Bruyeres-le-Chatel, France\\
$^{4}$Physik-Institut, University of Zurich, Winterthurerstrasse 190, 8057 Z\"urich, Switzerland\\
$^{5}$Dipartimento di Fisica e Astronomia, Universit\`{a} di Bologna, viale Berti Pichat 6/2, 40127, Bologna, Italy\\
$^{6}$Harvard-Smithsonian Center for Astrophysics, 60 Garden St., Cambridge, MA 02138, USA\\
$^{7}$Mullard Space Science Laboratory, University College London, Holmbury St Mary, Dorking, Surrey RH5 6NT, UK\\
$^{8}$Institute for Computational Science, University of Zurich, Winterthurerstrasse 190, 8057 Z\"urich, Switzerland}}

\begin{document}

\maketitle

\begin{abstract}
We measure the projected density profile, shape and alignment of the stellar and dark matter mass distribution in 11 strong-lens galaxies. We find that the projected dark matter density profile -- under the assumption of a Chabrier stellar initial mass function -- shows significant variation from galaxy to galaxy. Those with an outermost image beyond $\sim 10$\,kpc are very well fit by a projected NFW profile; those with images within 10\,kpc appear to be more concentrated than NFW, as expected if their dark haloes contract due to baryonic cooling. We find that over several half-light radii, the dark matter haloes of these lenses are rounder than their stellar mass distributions. While the haloes are never more elliptical than $e_{dm} = 0.2$, their stars can extend to $e_* > 0.2$. Galaxies with high dark matter ellipticity and weak external shear show strong alignment between light and dark; those with strong shear ($\gamma \gtrsim 0.1$) can be highly misaligned. This is reassuring since isolated misaligned galaxies are expected to be unstable. Our results provide a new constraint on galaxy formation models. For a given cosmology, these must explain the origin of both very round dark matter haloes and misaligned strong-lens systems.
\end{abstract}


\section{Introduction}\label{sec:introduction}

The ellipticity and shape of the stellar component relative to their host dark matter halo encodes information both about our current cosmological model $\Lambda$CDM and galaxy formation \citep[e.g.][]{1994ApJ...431..617D,2001ApJ...551..294I,2004ApJ...611L..73K,2007MNRAS.378...55M,2007arXiv0707.0737D,2012MNRAS.424L..16L,2014JPhG...41f3101R}. `Dark-matter-only' (DMO) simulations in $\Lambda$CDM predict dark matter haloes that are triaxial \citep{1991ApJ...378..496D,1992ApJ...399..405W,1996ApJ...462..563N,2002ApJ...574..538J}, with mean `shape parameter' $\langle q \rangle = (b+c)/2a \sim 0.8$ (where $a > b > c$ are the long, intermediate and short axes of the figure; \citealt{2007MNRAS.378...55M}). This corresponds to a typically {\it prolate} halo. However, including `baryons' (stars and gas) in the models produces haloes that are significantly rounder and -- at least for disc galaxies -- well-aligned with the light distribution \citep{1991ApJ...377..365K,1994ApJ...431..617D,2007arXiv0707.0737D}. Halo shapes and alignments also constrain alternative gravity models \citep{2001MNRAS.327..552M,2004ApJ...610L..97H,2005MNRAS.361..971R,2012PhRvD..86h3507F,2013MNRAS.434.2971D}. If stars dominate the mass of the galaxy, we expect the light and mass distribution to be highly correlated; if dark matter is present, however, such correlations can, at least in principle, be broken.

Strong lensing provides a unique probe of the alignment and shape of the total mass distribution in galaxies \citep[e.g.][]{1986ApJ...310..568B,1992grle.book.....S,1998ApJ...509..561K,2000ApJ...543..131K,2006ApJ...649..599K,2007AJ....134..668A,2008MNRAS.383..857F,2010ApJ...724..511A,2012MNRAS.424..104L}. For `red and dead' ellipticals that are largely devoid of gas, their baryonic content can be mapped through stellar population synthesis modelling of their light distribution alone \citep[e.g.][]{2005ApJ...623L...5F,2006ApJ...640..662T,2008MNRAS.383..857F}. Furthermore, such systems are dense enough to produce strong lensing effects, opening up the possibility of directly comparing the light and mass in these galaxies \citep{1998ApJ...509..561K,2008MNRAS.383..857F,2009ApJ...690..670T,2012A&A...538A..99S}. Previous work in the literature has found that the light and mass are well-aligned (though a mis-match of up to 10$^\circ$ is not uncommon; e.g. \citealt{2012A&A...538A..99S}). However, results on the ellipticity of light and mass agree less well, with \citet{2012A&A...538A..99S} finding a strong correlation and \citet{1998ApJ...509..561K} and \citet{2008MNRAS.383..857F} finding none. It is difficult, however, to compare the results between these different studies because they use different lens modelling techniques; different definitions of ellipticity; and different radii over which the shapes and alignments are probed. Furthermore, none to date have applied their methodology to mock data to determine the robustness of the results.

Weak lensing can also be used to probe the shape and alignment between the luminous and dark matter distributions within galaxies. However, this requires that the galaxies are `stacked' \citep{2000astro.ph..6281B,2000ApJ...538L.113N}, typically by aligning the major axes of the light distribution \citep[e.g.][]{2004ApJ...606...67H}. This will lead, in absence of contaminating effects \citep[e.g. intrinsic alignments;][]{2004PhRvD..70f3526H}, to an isotropic mean shear around galaxies if the mass distributions are spherical, or if dark and light are randomly aligned. The weak lensing measurement has been performed on data by various groups \citep[][]{2004ApJ...606...67H,2006MNRAS.370.1008M,2007ApJ...669...21P,2012A&A...545A..71V} with inconclusive results. The analyses differ mainly in the data sets used; the selection of lens galaxies; the estimators employed; and the treatment of systematics \citep[see e.g.][]{2015arXiv150704301S}. For example, \citet{2015arXiv150603536C} have recently made the measurement on Luminous Red Galaxies (LRG) taken from the Sloan Digital Sky Survey (SDSS) reporting a significant detection of dark matter halo ellipticity; while \citet{2015arXiv150704301S} performed a similar measurement on blue and red galaxies from the Canada France Hawaii Lensing Survey (CFHTLenS) finding no definite detection.

Recently, \citet{2014MNRAS.445.2181C} introduced a new non-parametric lens tool, \Glass. Applying this to a large suite of mock data, we showed that mass and light can only reliably be disentangled in strong lens systems if: i) there are at least four images; and ii) time delay data are available and/or the stellar mass contributes significantly to the potential. In this paper, we collate data of the above quality, compiling a sample of 11 strong lens galaxies. We apply \Glass\ to these lenses to non-parametrically measure the shape and alignment of the stars and {\it dark matter} in these lens galaxies, for the first time. This differs from previous works that have all compared the light distribution with the total mass, rather than the dark matter. Since the stellar component often dominates the central potential, the total mass naturally correlates with the light, potentially masking theoretically interesting results about the dark matter distribution. Our comparison between the stellar and dark matter components is made possible by the fact that \Glass\ uses the stellar mass distribution as a prior on the mass map, ensuring that the dark matter map is always positive.

This paper is organised as follows. In Section~\ref{sec:shapemethod}, we briefly review the \Glass\ code and define our method to assess shape and alignment of lens galaxies. In Section~\ref{sec:data}, we present our data compilation with references. In Section~\ref{sec:results}, we present our results. We discuss the implications of these results in Section~\ref{sec:discussion}. Finally, we conclude in Section~\ref{sec:conclusions}.

Throughout this work we assume a flat $\Lambda$CDM model with matter density $\Omega_{m} = 0.28$, dark energy density $\Omega_{\Lambda} = 0.72$, and inverse Hubble constant $H_{0}^{-1} = 13.7$ Gyr.

\section{The lens models}\label{sec:shapemethod}
\setlength\tabcolsep{7pt}
\begin{table*}
  \begin{center}
    \begin{tabular}{l l l l l l l l}
      \multirow{2}{*}{Lens} & \multirow{2}{*}{$z_{L}$} & \multirow{2}{*}{$z_{S}$} & $\Delta\phi$ & \multirow{2}{*}{$R_{L}/R_{e}$} & $M_s(<2R_e)$ & \multirow{2}{*}{Env} & \multirow{2}{*}{$\theta_{Env}$ [$^{\circ}$]} \\ 
       & & & [kpc] & & [$10^{10}M_{\odot}$] & & \\ \hline
      0047 & 0.485 & 3.60 & 12.82 & $1.45\pm0.04$ & $11.58\pm0.43$ & G(9)$^{a}$ & ... \\
      0414 & 0.960 & 2.64 & 16.01 & $1.85\pm0.05$ & $19.90\pm2.29$ & ... & ... \\
      0712 & 0.410 & 1.34 & 6.82  & $1.15\pm0.03$ & $5.46\pm0.52$ & ... & ... \\
      0911 & 0.769 & 2.8  & 23.16 & $3.09\pm0.05$ & $14.52\pm1.87$ & C$^{b}$ & -160$^{b}$ \\
      0957 & 0.356 & 1.41 & 29.98 & $3.51\pm0.04$ & $20.92\pm0.96$ & C$^{c}$ & 60$^{f}$ \\
      1115 & 0.310 & 1.72 & 10.76 & $2.86\pm0.06$ & $5.61\pm1.19$ & G(13)$^{a}$ & -125$^{g}$ \\
      1422 & 0.337 & 3.62 & 6.02  & $4.49\pm0.06$ & $2.91\pm0.45$ & G(17)$^{a}$ & 145$^{g}$ \\
      1608 & 0.630 & 1.39 & 13.92 & $1.82\pm0.01$ & $27.99\pm1.63$ & G(8)$^{d}$ & 30 resp. -5$^{d,h}$\\
      2016 & 1.010 & 3.3  & 26.22 & $6.12\pm0.14$ & $6.34\pm1.99$ & C(69)$^{e}$ & 130 - 140$^{e}$ \\
      2045 & 0.870 & 1.28 & 14.46 & $1.48\pm0.03$ & $14.05\pm1.03$ & ... & ... \\
      2237 & 0.039 & 1.7  & 1.40  & $0.89\pm0.01$ & $1.15\pm0.12$ & ... & ... \\
    \end{tabular}
    \caption[width=\linewidth]{The most relevant lens properties for this work are listed \citep[for an expanded version of this table see][]{2011ApJ...740...97L}. References can be found in Section~\ref{sec:data}. The columns are from left to right: the lens redshift $z_L$; the redshift of the source $z_S$; the maximum angular separation between two images $\Delta\theta$; the ratio of the radius of the outermost image $R_L$ and the effective Petrosian half-light radius of the lens $R_e$; the stellar mass $M_s$ within $2R_e$; information on the environment of the lens; and the position angle to the centroid of the corresponding group/cluster (measured north through east). C and G denote a known cluster or group environment, respectively. The number in the brackets is the number of confirmed members. Ellipses indicate lens galaxies with no group members identified so far. \newline $^{a}$ \citet{2011ApJ...726...84W}; $^{b}$ \citet{2001ApJ...555....1M}; $^{c}$ e.g. \citet{1992MNRAS.254P..27G}; $^{d}$ \citet{2006ApJ...642...30F}; $^{e}$ \citet{2003MNRAS.344..337T}; $^{f}$ \citet{2002ApJ...565...96C}; $^{g}$ \citet{2004ApJ...610..686G}. \newline $^{h}$ The position angle is given for the unweighted resp. the luminosity-weighted centroid estimation.}
    \label{tab:lensproperties}
  \end{center}
\end{table*}

While it is possible to compute shape parameters for a lensing galaxy by fitting a parametric shape to the lens, it makes the definition of a shape estimate dependent on the parametric form being used. Moreover, the commonly-used parametric forms for modelling lensing galaxies \citep[e.g.][]{2001astro.ph..2341K} do not allow for features like twisting isodensity contours, which can arise from the projection of triaxial ellipsoids with no intrinsic twists \citep[e.g.][and references therein]{1978ComAp...8...27B}. To avoid these problems, we use non-parametric ellipticity estimators, defined as follows.

Lenses are modelled as free-form mass distributions, consisting of mass tiles or pixels. Such lens models are also called non-parametric, which really just means that many more parameters than data constraints are used. Thus the lens models are non-unique, and hence we build {\it ensembles} of these free-form mass distributions. From such a mass map, an inertia tensor is defined as:
\begin{equation}\label{eq:inertiatensor}
I = 
\begin{pmatrix}
 \sum_\theta M(\boldsymbol{\theta})\theta^{2}_{y} & -\sum_\theta M(\boldsymbol{\theta})\theta_{x}\theta_{y} \\
-\sum_\theta M(\boldsymbol{\theta})\theta_{x}\theta_{y} & \sum_\theta M(\boldsymbol{\theta})\theta^{2}_{x}
\end{pmatrix}
\end{equation}
where the sum is over mass pixels, and $M(\boldsymbol{\theta})$ is the mass in a pixel. The eigenvectors of this inertia tensor give the ellipticity axes, and our ellipticity estimator is
\begin{equation}\label{eq:shapeestimate}
    e \equiv \frac{\lambda_{1}-\lambda_{2}}{\lambda_{1}+\lambda_{2}},
\end{equation}
where $\lambda_{1}$ and $\lambda_{2}$ ($\lambda_{1} \geq \lambda_{2}$) are the eigenvalues. We use this ellipticity estimator to quantify the shapes both of the stellar-mass distribution $e_*$ and of the dark-matter distribution $e_{dm}$. By default, we use the dark and stellar mass distributions out to the outermost lensing image. In Figure \ref{fig:wedgesradii}, we explore the effect of averaging instead over multiples of the half-light radius $R_e$. The mass pixels used in this work are larger than image pixels, but at most 4\% of the diameter of the the full mass map. The central region has smaller pixels, to allow for a density cusp at the centre.

We estimate the orientation of the distributions of luminous and dark matter by computing the angles $\theta_{*}$ and $\theta_{dm}$ of the eigenvector corresponding to the eigenvalue $\lambda_{1}$ ($\lambda_{1} \geq \lambda_{2}$) relative to the $x$-axis. The misalignment angle $\Delta\theta$ between the distributions is then defined as:
\begin{equation}\label{eq:misalignment}
  \Delta\theta = \theta_{dm} - \theta_{*}.
\end{equation}

Each model ensemble consists of $10^4$ models. We apply our shape and alignment estimators to each model, in this way obtaining the median and 68\% confidence intervals of $e$ and $\Delta \theta$, for each lens.

The non-parametric mass models themselves are constructed using the \Glass\ framework for modelling multiple-image lenses \citep{2014MNRAS.445.2181C}. \Glass\ produces ensembles of mass maps, each of which reproduces exactly the observed image positions, image parities and time delays (if measured). The lensing data must be in the form of point images, which could be quasars or simply point-like features in otherwise extended images, but this was not a problem for the lenses studied in this work. The estimated stellar mass (explained below in Section~\ref{sec:data}) is taken as a lower limit on the total mass. The additional mass that needs to be put on top, in order to reproduce the lensing observables, is interpreted as dark matter. The lensing observables and assumed non-negative dark matter by themselves leave the mass distribution under-determined. Hence, an additional prior is used, requiring the mass distribution to be centrally concentrated, with the same centre as the galaxy light (but not necessarily the same shape). The precise formulation of the prior is given in Section~3 of \citet{2014MNRAS.445.2181C} \citep[alternatively, for a more intuitive account of \Glass\ see Section~3.2 of][]{2015MNRAS.447.2170K}.

\Glass\ is closely related to an earlier code \PixeLens\ \citep{2004AJ....127.2604S,2008ApJ...679...17C}, but has a completely different code base. An important improvement is a better sampling algorithm for high dimensional spaces \citep{2012MNRAS.425.3077L} that eliminates the excessive weight \PixeLens\ tended to give to extreme models.

Most relevant to the present work are Section~5.2 and Figure~8 of \citet{2014MNRAS.445.2181C}, showing the recovery of $\lambda_2/\lambda_1$. The typical errors in this quantity are 10--20\%.

\section{Data}\label{sec:data}
Stellar mass maps are taken from \citet{2011ApJ...740...97L}. In that work, the surface brightness distribution of the galaxy images in all available HST bands was first fitted using {\sc Galfit v2.03b} \citep{2002AJ....124..266P}. This was then converted to a stellar mass map using the stellar-population synthesis (SPS) models of \citet{2003MNRAS.344.1000B}. The population synthesis is marginalised over the star formation epoch and time scales and over the stellar metallicity. The initial mass function (IMF) was taken to be the log-normal form given by \citet{2003PASP..115..763C}; a more bottom-heavy IMF would change the mass normalisation \citep[cf.][]{2014ApJ...793...96S} but not the shape of the inferred stellar mass distribution \citep[unless the IMF presents significant intrinsic deviations locally, see e.g.][]{2015MNRAS.447.1033M}. We note that the effect of recent claims towards a non-universal IMF on stellar mass-to-light ratio (M/L) is still under debate. While for massive elliptical galaxies, dynamical studies seem to point at contributions at the level of the Salpeter IMF or slightly heavier \citep{2013MNRAS.432.1862C}, spectral line strength constraints can still accommodate a wider range of the stellar M/L, depending on the functional form of the IMF \citep{2013MNRAS.429L..15F}. For at least one lens in our sample, \textit{Q2237+030}, the lensing estimates on the mass distribution are inconsistent with a Salpeter IMF \citep{2010MNRAS.409L..30F}. We additionally tested varying the IMF for one lens ({\it B1422+231}), finding that a Salpeter IMF is too heavy and gives no solutions. We will perform a more systematic study of the IMF and varying stellar M/L in a forthcoming publication.

The original sample we consider, is presented in \citet{2011ApJ...740...97L} and consists of 21 lens systems. It is a subsample of the CfA-Arizona Space Telescope LEns Survey\footnote{http://www.cfa.harvard.edu/castles/} (CASTLES) sample. As shown in \citet{2014MNRAS.445.2181C}, the recovery of the shape information of the lens is only good if there are at least four images. Therefore, we only consider the quad lenses and the twin double \textit{Q0957+561}. This yields a sample consisting of 11 lenses. We discuss each of these systems, next. We note that all the position angles are measured north through east and relative to the lens galaxy.

\subsection{The individual lenses}
\setlength\tabcolsep{5pt}
\begin{table}
  \begin{center} \begin{tabular}{l l l l}
      Lens & Time delays [days] & Symmetric? & Comments \\ \hline
      0047 & ... & ... & ... \\
      0414 & ... & ... & ... \\
      0712 & ... & ... & ... \\
      0911 & $\Delta t_{BA}=146^{+4}_{-4}$ & yes & ... \\
      0957 & $\Delta t_{BA}=416.5^{+1}_{-1}$ & yes & ... \\
      1115 & $\Delta t_{BA}=12.0^{+2}_{-2}$ & ... & ... \\
           & $\Delta t_{DC}=4.4^{+3.2}_{-2.4}$ & ... & ... \\
      1422 & ... & yes & ... \\
      1608 & $\Delta t_{BA}=31.5^{+2}_{-1}$ & ... & ... \\
           & $\Delta t_{CA}=36.0^{+1.5}_{-1.5}$ & \\
           & $\Delta t_{DA}=77.0^{+2}_{-1}$ & \\
      2016 & ... & ... & Smaller pixels \\
      2045 & ... & yes & $\gamma\leq 0.1$ \\
      2237 & ... & ... & ... \\
    \end{tabular}
    \caption[width=\linewidth]{Lenses where additional data or priors were used (cf.~Section~\ref{sec:shapemethod}). $\Delta t_{BA}$ and so on denote time delays between images as labelled in Table~\ref{tab:lenspositions}; $\gamma$ refers to the allowed external shear; Symmetric `yes' means the mass map is symmetric under rotation by $180^\circ$; `Smaller pixels' indicates that the mass pixels are 3\% (rather than 4\%) of the diameter of the whole mass map.}
    \label{tab:lenspriors}
  \end{center}
\end{table}
\setlength\tabcolsep{6pt}

\textit{Q0047-2808} (hereafter \textit{0047}) is a luminous early-type galaxy \citep{1996MNRAS.278..139W}. It is part of a group of 9 members, all of which are spectroscopically confirmed \citep{2011ApJ...726...84W}.

\textit{MG0414+0534} (hereafter \textit{0414}) is a passively evolving early-type galaxy \citep{1999AJ....117.2034T}. A close luminous satellite galaxy was found north-west of the lens \citep{1993AJ....105....1S}. It, however, is more likely a foreground object with a redshift of ~0.38 \citep{2011MNRAS.413L..86C}.

\textit{B0712+472} (hereafter \textit{0712}) is an early-type galaxy \citep{1998MNRAS.296..483J,1998AJ....115..377F}. There is a foreground group of about 10 member galaxies at $z\sim0.3$ \citep{2002AJ....123..627F}.

\textit{RXJ0911+0551} (hereafter \textit{0911}) is an early-type galaxy \citep{1997A&A...317L..13B,2012A&A...538A..99S}, and has measured time delays \citep{2002ApJ...572L..11H}. It lies on the outskirts of a cluster of which X-ray emission can be detected yielding a temperature of 2.3 keV \citep{2001ApJ...555....1M}. The center of the cluster lies at a position angle of about -160$^{\circ}$. There is furthermore a satellite galaxy to the north-west \citep{2000ApJ...544L..35K}.

\textit{Q0957+561} (hereafter \textit{0957}) is a cD galaxy lying close to the centre of a cluster with a high spiral galaxy-fraction \citep[e.g.][]{1992MNRAS.254P..27G,1994A&A...291..411A,1998ApJ...504..661C}. Due to the large image separation, large physical scales are probed. Identified X-ray emission furthermore locates the center of the cluster at a position angle of about 60$^{\circ}$, close to the lens galaxy \citep{2002ApJ...565...96C}. The lens has a measured time delay \citep[e.g.][]{2012A&A...540A.132S}. They however also find a three-day lag between the g- and r-bands, a disagreement at the 2$\sigma$-level. They argue that this effect can be accounted for by the presence of a substructure and chromatic dispersion. We find that the results do not change significantly for either estimate and choose therefore the g-band measurement.

\textit{PG1115+080} (hereafter \textit{1115}) is an early-type galaxy \citep{1980Natur.285..641W,2005ApJ...626...51Y}. It has measured time delays \citep[see e.g.][]{1997ApJ...475L..85S}. In this work, we use recent estimates of the time delays by \citet{2010MNRAS.406.2764T} that differ from previously found values by e.g. \citet{1997ApJ...489...21B}. The environment of the lens was analysed thoroughly \citep{2006ApJ...641..169M,2011ApJ...726...84W}. It is part of a small group of 13 members. Also, X-ray emission was detected from the corresponding group at a position angle of about -125$^{\circ}$ and it yields a temperature of 0.8 keV \citep{2004ApJ...610..686G}.

\textit{B1422+231} (hereafter \textit{1422}) is an early-type galaxy \citep{1992MNRAS.259P...1P,1994AJ....107...28Y}. Although time delays have been reported \citep{2001MNRAS.326.1403P}, it is possible that the measurements are aliases of the actual values \citep{2003AJ....126...29R} and hence are not included in our analysis. The lens is part of a group with 16 spectroscopically confirmed member galaxies \citep{2006ApJ...641..169M}. Using newer data, additional members could be identified \citep{2011ApJ...726...84W}. \citet{2004ApJ...610..686G} also detect X-ray emission from the corresponding group originating from a position angle of about 145$^{\circ}$ at a temperature of 1.0 keV.

\textit{B1608+656} (hereafter \textit{1608}) was first reported by \citet{1995ApJ...447L...5M}. It consists of two merging galaxies. The main galaxy is an early-type galaxy which is disrupted by a smaller, probably late-type galaxy \citep{2003ApJ...584..100S}. The system has measured time delays \citep{2002ApJ...581..823F}. The environment has been analysed and a group of 8 (9 if the merging galaxies are counted individually) member galaxies was identified \citep{2006ApJ...642...30F}. Furthermore, the center was identified to be at a position angle of about 30$^{\circ}$ (resp. -5$^{\circ}$) if the member galaxies' positions are averaged without weighting (resp. weighted by luminosity). However, no significant X-ray emission was detected from the surrounding group \citep{2005ApJ...625..633D}. The data seem to indicate four other groups along the line-of-sight.

\textit{MG2016+112} (hereafter \textit{2016}) is a giant elliptical galaxy \citep{1984Sci...223...46L,1986AJ.....91..991S}. It is the farthest lens we consider in this sample. It lies in a cluster that consists of 69 probable galaxies \citep{2003MNRAS.344..337T}. The cluster shows a high density of galaxies close to the lens in a south-east direction, at about $\sim130-140^{\circ}$. Two lensed images are very close in this fold lens, but could be resolved by \citet{2009MNRAS.394..174M}.

\textit{B2045+265} (hereafter \textit{2045}) is probably an elliptical galaxy \citep{2007MNRAS.378..109M}. It was initially classified as a late-type Sa galaxy \citep{1999AJ....117..658F}, however the velocity dispersion seems too high. As the source redshift is rather low, a large lens mass is required. \citet{2007MNRAS.378..109M} therefore conclude that it is more likely an elliptical galaxy. North-west of the lens there is evidence for a potential group at a similar redshift as the lens \citep{1999AJ....117..658F}. There is furthermore evidence for a dwarf satellite galaxy within the Einstein radius of the lens system \citep{2007MNRAS.378..109M}. As the measurements are however inconclusive, we choose to not include this satellite in our models.

\textit{Q2237+030} (hereafter \textit{2237}) is a barred spiral \citep{1988AJ.....95.1331Y}. With a redshift of just $z\sim0.04$ it is the closest lens system of the sample. Due to its low redshift, the probed physical scales are small, making the bulge component of the galaxy dominant. This makes the treatment of the stellar populations easier, as derivations of the stellar M/L are more difficult in the presence of star forming regions typical of disc galaxies, where there are no simple stellar populations usually assumed in SPS models.

Further information on the sample can be found in \citet{2011ApJ...740...97L} and \citet{2012A&A...538A..99S}. The most important quantities of each system are given in Table~\ref{tab:lensproperties}. Due to the different angular separations of the images and the different angular diameter distances, in each lens galaxy we necessarily probe slightly different scales.

\section{Results}\label{sec:results}
\begin{figure*}
  \centering
  \includegraphics[width=.7\linewidth]{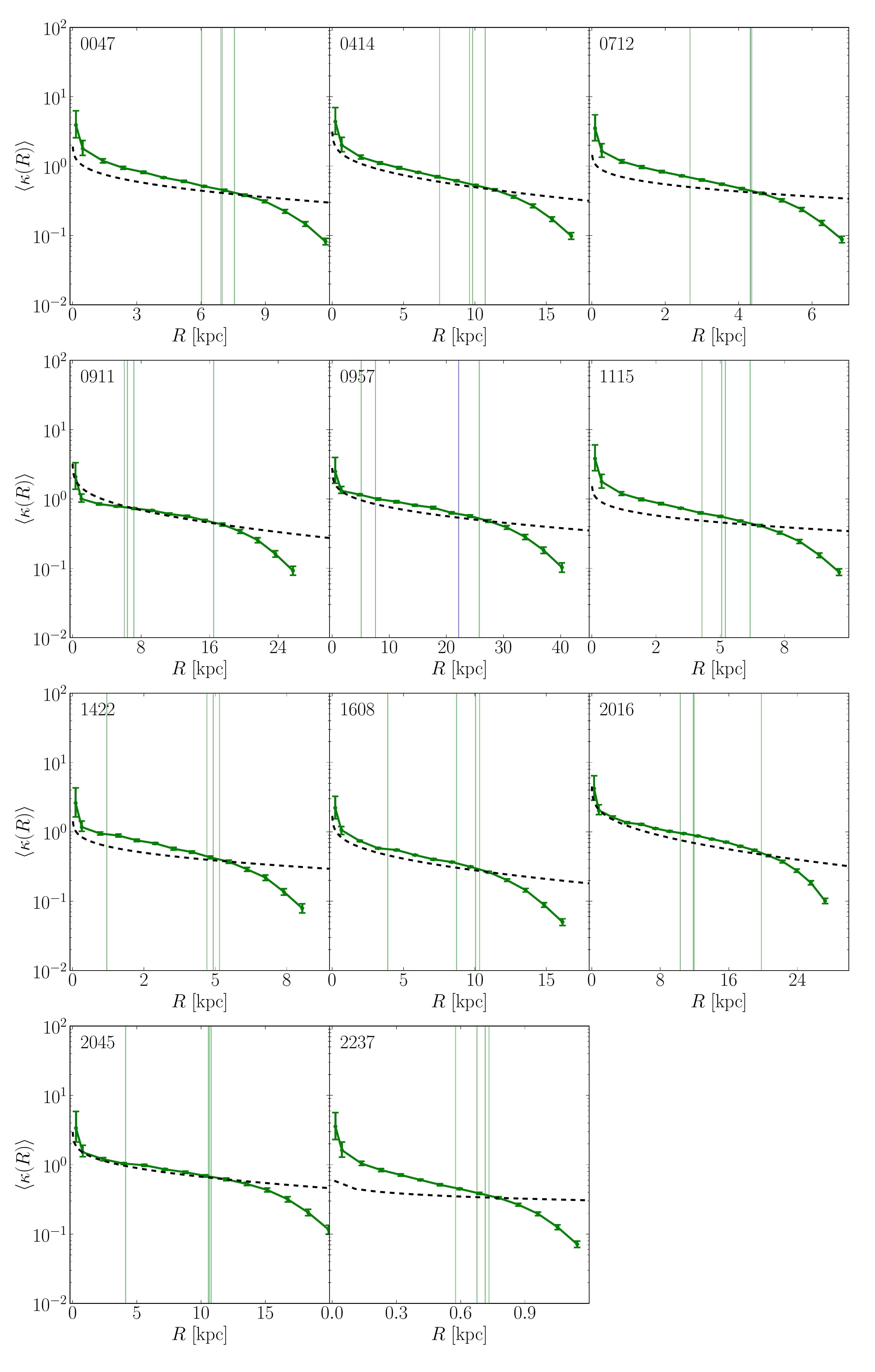}
  \caption[width=\linewidth]{Azimuthally averaged convergence $\langle\kappa\rangle$ within radii $R$ of the dark matter distribution. The vertical lines denote the radial distances of the source images from the centre of the lens galaxies. For \textit{0957}, there are two lensed components of the source galaxy denoted by the green and blue vertical lines. The green curves show the contributions by the dark matter in the strong lens galaxies as reconstructed by \Glass. The dashed black lines show the corresponding profiles for dark matter haloes that follow a NFW density profile. The parameters in the NFW distribution are calibrated such that the convergence $\langle\kappa\rangle$ matches the corresponding values of the reconstructed profiles at one pixel past the last image radius. Notice that those lenses where the outermost image is within $\sim 10$\,kpc appear contracted as compared to an NFW profile; those where the outermost image is further out $> 10$\,kpc are well fit by the NFW form. In all cases there is a sharp fall off in the \Glass\ dark matter profile at large radii due to the edge of the mass map.}
  \label{fig:kappaplot}
\end{figure*}

\begin{figure*}
  \centering
  \includegraphics[width=\linewidth]{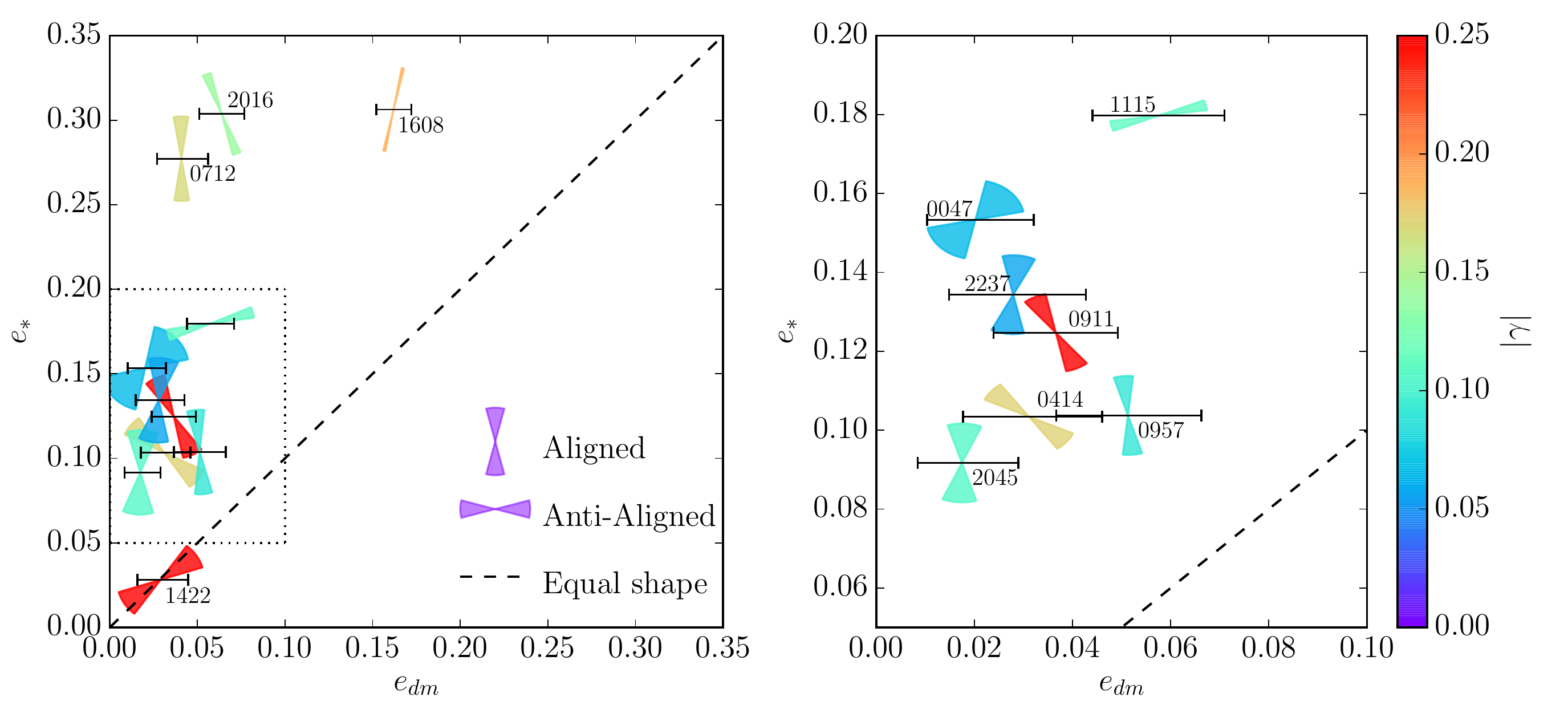}
  \caption[width=\linewidth]{The reconstructed lens galaxies are displayed according to the reconstructed total shapes of the dark matter haloes $e_{dm}$ and the stellar components $e_{*}$. The wedges show the alignment of the semi-major axes of the components. In case of alignment of the respective semi-major axes (Aligned), the wedges are vertical. In case of alignment of the semi-major with the semi-minor axis (Anti-Aligned), the wedges are horizontal. The opening angle of each wedge displays the statistical error in the shape estimate of the dark matter haloes. Both the uncertainties in the alignment and the ellipticity of the dark matter distribution denote the ranges $68\%$ of all the mass models of the ensemble of solutions lie within. The lens galaxies are additionally color-coded in terms of the required external shear in reconstructing the mass distribution. The right panel shows the region in the left panel denoted by a dashed rectangle in greater detail.}
  \label{fig:wedgesall}
\end{figure*}

\begin{table*}
  \begin{center}
    \begin{tabular}{l l l l l l l l l l l l}
      \multirow{2}{*}{Lens} & \multirow{2}{*}{Env} & \multirow{2}{*}{$\theta_{Env}$ [$^{\circ}$]} & $r_{vir}$ & $M_{vir}$ & \multirow{2}{*}{$|\gamma|$} & \multirow{2}{*}{$\theta_{g}$ [$^{\circ}$]} & \multirow{2}{*}{$\Delta\theta$ [$^{\circ}$]} & \multirow{2}{*}{$e_{*}$} & \multirow{2}{*}{$e_{dm}$} & \multirow{2}{*}{$e_{dm}/e_{*}$} & \multirow{2}{*}{$f_h$} \\
       & & & [kpc] & $[10^{12} \ M_{\odot}]$ & & & & & & & \\ \hline
      0047 & G(9) & ... & 360.7 & 10.0 & 0.07$\pm$0.01 & 120$\pm$5 & -41$\pm$33 & 0.15 & 0.02$\pm$0.01 & 0.13$\pm$0.07 & 0.02$\pm$0.15 \\
      0414 & ... & ... & 256.9 & 12.4 & 0.17$\pm$0.02 & 166$\pm$4 & 50$\pm$15 & 0.10 & 0.03$\pm$0.01 & 0.30$\pm$0.14 & -0.05$\pm$0.15 \\
      0712 & ... & ... & 446.5 & 15.5 & 0.17$\pm$0.01 & 141$\pm$3 & -0$\pm$10 & 0.28 & 0.04$\pm$0.01 & 0.15$\pm$0.05 & 0.15$\pm$0.05 \\
      0911 & C & -160 & 369.1 & 22.7 & 0.28$\pm$0.02 & 99$\pm$2 & 25$\pm$14 & 0.12 & 0.04$\pm$0.01 & 0.29$\pm$0.10 & 0.19$\pm$0.12 \\
      0957 & C & 59 & 1053.1 & 175.2 & 0.09$\pm$0.03 & 159$\pm$8 & 4$\pm$11 & 0.10 & 0.05$\pm$0.01 & 0.49$\pm$0.14 & 0.49$\pm$0.14 \\
      1115 & G(13) & -127 & 616.8 & 31.1 & 0.11$\pm$0.01 & 154$\pm$3 & -75$\pm$7 & 0.18 & 0.06$\pm$0.01 & 0.32$\pm$0.07 & -0.28$\pm$0.08 \\
      1422 & G(17) & 147 & 450.8 & 13.0 & 0.26$\pm$0.01 & 39$\pm$1 & -51$\pm$18 & 0.03 & 0.03$\pm$0.01 & 1.03$\pm$0.51 & -0.23$\pm$0.65 \\
      1608 & G(8) & 30 resp. -5$^{a}$ & 317.1 & 10.0 & 0.19$\pm$0.01 & 145$\pm$2 & -12$\pm$1 & 0.31 & 0.16$\pm$0.01 & 0.53$\pm$0.03 & 0.48$\pm$0.03 \\
      2016 & C(69) & $130-140$ & 299.3 & 22.1 & 0.14$\pm$0.02 & 42$\pm$4 & 20$\pm$6 & 0.30 & 0.06$\pm$0.01 & 0.21$\pm$0.04 & 0.16$\pm$0.04 \\
      2045 & ... & -60 & 562.8 & 104.0 & 0.11$\pm$0.02 & 23$\pm$7 & -4$\pm$20 & 0.09 & 0.02$\pm$0.01 & 0.19$\pm$0.11 & 0.19$\pm$0.11 \\
      2237 & ... & ... & 1961.0 & 485.3 & 0.06$\pm$0.01 & 159$\pm$4 & -7$\pm$19 & 0.13 & 0.03$\pm$0.01 & 0.21$\pm$0.10 & 0.20$\pm$0.11 \\
    \end{tabular}
    \caption{Fitted virial masses and radii and the ellipticity values of the stellar and dark matter components of the reconstructed strong lens galaxies and their corresponding ratios. `Env' and $\theta_{Env}$ are the `Environment' resp. $\theta_{Env}$ columns in Table~\ref{tab:lensproperties}; $r_{vir}$ and $M_{vir}$ are the virial radii and masses of the fitted NFW-profiles (see Section~\ref{sec:radialprofiles}); $|\gamma|$ and $\theta_{g}$ denote magnitude and induced position angle (measured north through east) of the required external shear in the modeling of the strong lens galaxy; the ellipticities $e_{*}$ and $e_{dm}$ are defined by Eq.~\ref{eq:shapeestimate}; $\Delta\theta=\theta_{dm}-\theta_{*}$ refers to the misalignment angle between the distributions of luminous and dark matter; $e_{dm}/e_{*}$ denotes the ratio of the ellipticities of the dark matter relative to the stellar distribution; and $f_h=(e_{dm}/e_{*})\cdot\mathrm{cos} 2\Delta\theta$ refers to the ratio of the ellipticity component of the dark matter distribution projected along the stellar distribution.}
    \label{tab:ellipratios}
  \end{center}
\end{table*}

\begin{figure*}
  \centering
  \includegraphics[width=.75\linewidth]{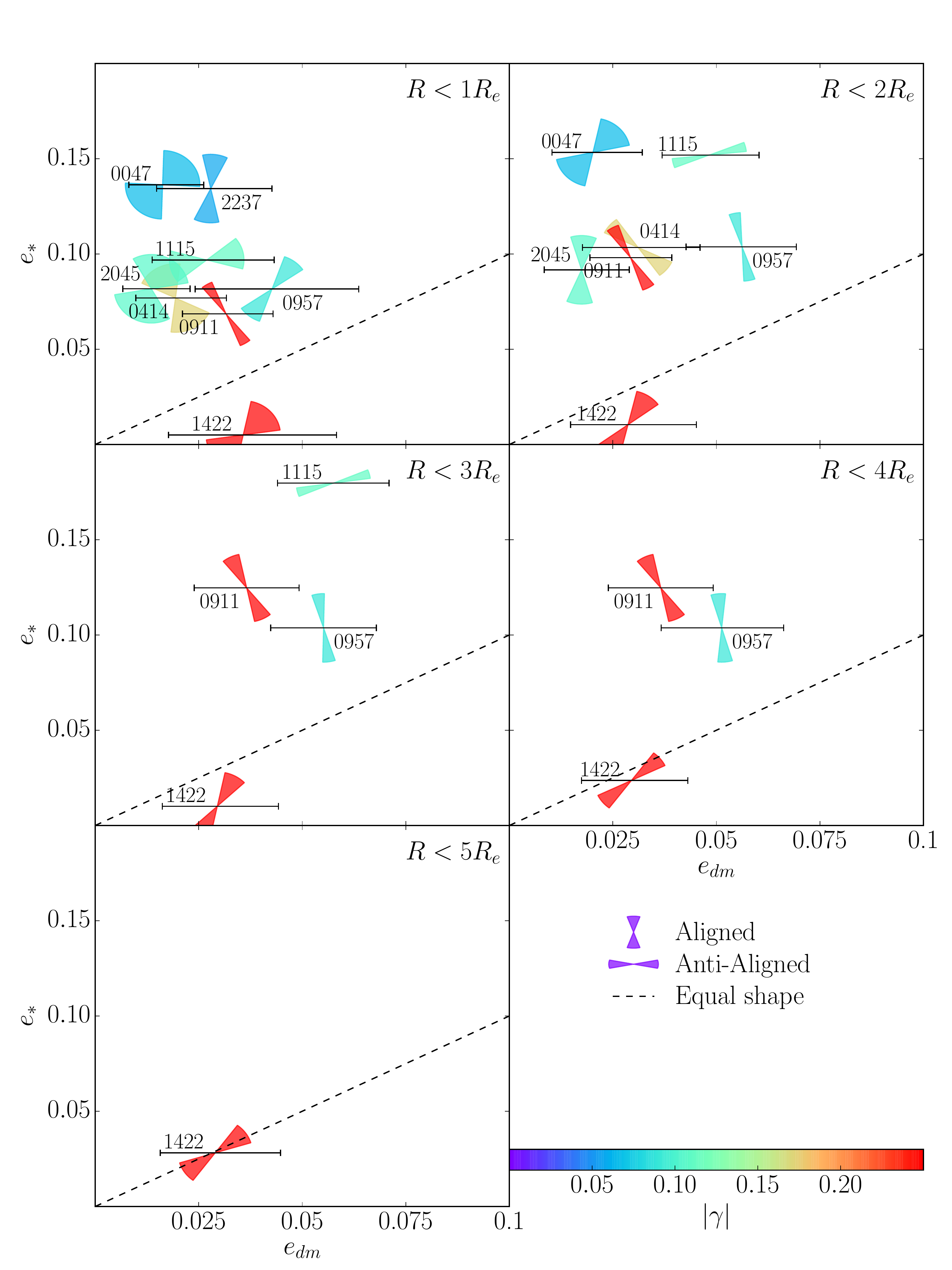}
  \caption[width=\linewidth]{Similar to Figure~\ref{fig:wedgesall}. In each of the subfigures however we only consider pixels within 1, 2, 3, 4, or 5 $R_e$. If the outermost image of a system lies within one of the limiting radii, the system is kept in this panel, but dropped from subsequent panels.}
  \label{fig:wedgesradii}
\end{figure*}

We have applied \Glass\ to the data set described in Section~\ref{sec:data} consisting of 11 lenses. The reconstructed stellar and dark mass maps for each individual lens are given in the Appendix.

\subsection{Radial dark matter profiles}\label{sec:radialprofiles}
Before discussing the shapes and alignments, we first measure the dark matter profiles of the strong lens galaxies, comparing these with Navarro-Frenk-White profiles \citep[NFW;][]{1996ApJ...462..563N} given by \citep{2011ApJ...740..102K}:
\begin{equation}\label{eq:nfw}
   \frac{\rho(r)}{\rho_{crit}} = \frac{\delta_{c}}{(cr/r_{vir})(1+cr/r_{vir})^2},
\end{equation}
where the critical density of the universe $\rho_{crit}(z)$ and $\delta_{c}(z)$ are parameters set by the cosmology considered, $r$ is the radius from the centre, and $c$ is the concentration parameter. The concentration is well-correlated with $M_{vir}$ with some redshift dependence \citep{2011ApJ...740..102K}:
\begin{equation}\label{eq:concentration}
 \begin{split}
   c(M_{vir}, z) = & c_{0}(z)\left(\frac{M_{vir}}{10^{12}h^{-1}M_{\odot}}\right)^{-0.075} \\ & \times \left[1+\left(\frac{M_{vir}}{M_{0}(z)}\right)^{0.26}\right],
 \end{split}
\end{equation}
where $c_{0}(z)$ and $M_{0}(z)$ are fitting parameters \citep[see Table 3 of][]{2011ApJ...740..102K}. For each lens galaxy we choose the parameters at the redshift closest to the lens redshift.

For a circular NFW profile truncated at the virial radius $r_{vir}$, the projected surface mass density $\Sigma(x)$, where $x = cr/r_{vir}$, can be expressed as \citep{2009JCAP...01..015B}:
\begin{equation}
 \begin{aligned}
  \Sigma(x) = & \frac{M_1}{r_{vir}^2}\frac{c^4}{2\pi(c^2+1)^2}\left\{\frac{c^2+1}{x^2-1}[1-F(x)]+2F(x) \right. \\
  & \left. -\frac{\pi}{\sqrt{c^2+x^2}}+\frac{c^2-1}{c\sqrt{c^2+x^2}}L(x)\right\},
 \end{aligned}
\end{equation}
where
\begin{equation}
  M_{vir} = M_{1} \frac{\tau^2}{(\tau^2+1)^2}\left[(\tau^2-1)\mathrm{ln}\tau+\tau\pi-(\tau^2+1)\right]
\end{equation}
and
\begin{equation}
  F(x) = \frac{\mathrm{cos}^{-1}(1/x)}{\sqrt{x^2-1}} \ \ \mathrm{and} \ \ L(x) = \mathrm{ln}\left(\frac{x}{\sqrt{c^2+x^2}+c}\right).
\end{equation}
The convergence $\kappa(x)=\Sigma(x)/\Sigma_{crit}$ can then be computed, where $\Sigma_{crit}$ is the critical surface density at the lens' redshift. Finally, $\langle\kappa(R)\rangle$ is obtained by azimuthally averaging.

As the virial radius $r_{vir}$ is related to the virial mass $M_{vir}$ and the concentration parameter $c$ is well-correlated with $M_{vir}$ (Eq.~\ref{eq:concentration}), there is only one degree of freedom, $M_{vir}$. We calibrate the mass in order for the derived convergence profiles $\langle\kappa(R)\rangle$ of the reconstructed dark matter distribution and the NFW-profile to match at one pixel past the last image radius.

Figure~\ref{fig:kappaplot} displays the results, while the virial masses (derived following the above procedure) are given in Table~\ref{tab:ellipratios}. In all cases the dark matter distribution either follows the NFW profile or is steeper, as expected if dark haloes contract in response to baryon cooling and star formation \citep[e.g.][]{1991ApJ...377..365K,1994ApJ...431..617D,2007arXiv0707.0737D}. At large radii, there is in all cases a sharp fall off in the \Glass\ dark matter density profiles. This owes to the edge of the mass map and for this reason we should not trust the reconstructed profiles outside of the outermost image. Interestingly, there is significant scatter from galaxy to galaxy in the shape of the azimuthally averaged dark matter density profile. This appears to correlate well with the physical radii probed. The galaxies that most closely follow an NFW profile -- \textit{0911}, \textit{0957}, \textit{0414}, \textit{1608}, \textit{2016}, \textit{2045} -- all have an outermost image at radii $> 10$\,kpc. The other lenses -- \textit{0047}, \textit{0712}, \textit{1115}, \textit{1422}, \textit{2237} -- probe smaller radii and show signs of contraction with respect to NFW. In the most extreme example -- \textit{2237} -- the outermost image probes only the bulge region of this galaxy ($\sim 0.8$\,kpc). The dark matter profile is substantially steeper than NFW while the derived virial mass is implausibly large (since we have not accounted for any such contraction in deriving it; see Table \ref{tab:ellipratios}). Our results suggest that the dark matter haloes in these lensing galaxies are broadly consistent with contracted NFW profiles. Lenses with close-in images -- that are expected to inhabit less massive dark matter haloes -- show more apparent contraction. There could be several explanations for this behaviour. It could be a selection effect where lower mass galaxies can only strong lens if they are substantially contracted; it could owe to more massive haloes being less affected by baryons; or it could be that we simply cannot yet resolve the contraction at the centre of the more massive systems. We will explore such questions in a forthcoming paper where we will present the dark matter radial profiles of a larger sample of lenses, considering varying stellar initial mass function, and explicitly modelling dark matter contraction.

\subsection{Shape and alignment of the dark matter haloes}
Figure~\ref{fig:wedgesall} shows our results for the shapes (see Section~\ref{sec:shapemethod}) and alignments of the dark matter and stellar mass distributions, while the values are listed in Table~\ref{tab:ellipratios}. In this Figure, the full distribution up to one pixel past the last-image radius $R_{L}$ is probed. There are several key points to note. Firstly, the reconstructed dark matter distribution of all lenses considered here is rounder than the stellar distribution. Secondly, most lenses lie in the bottom-left corner: the dark matter distributions have typically low eccentricities, while the stellar distributions can be rather elliptical. Thirdly, 3 out of the 11 systems, {\it0712}, {\it2016}, and {\it1608}, stand out due to their larger stellar ellipticity ($e_* > 0.25$). At the same time, they show a very strong alignment between their stellar and dark matter distributions. One of these lenses ({\it1608}) is a known merger of an early-type galaxy with a probable late-type galaxy \citep{2003ApJ...584..100S}. We speculate that the other two systems, {\it0712} and {\it2016}, may also be post-merger systems, explaining their similarities with {\it1608}.

Fourthly, notice that all dark matter and stellar distributions with weak external shear ($\gamma < 0.1$) are aligned apart from {\it0047} which is almost spherical ($e_{dm}\sim0.02$). By contrast, systems requiring a larger external shear ($\gamma > 0.1$; {\it0414}, {\it0911}, and {\it1422}) can display a sizeable misalignment between the distributions. The latter two systems are located in dense environments (see Table~\ref{tab:lensproperties}) that are likely responsible for these misalignments. As described furthermore in Section~\ref{sec:data}, {\it1422} is a challenging system to model the stellar distribution. For {\it0414} there is another galaxy along the line-of-sight contributing to the lensing effect \citep{2011MNRAS.413L..86C}. Besides these three lens systems, one stands out from the rest, however. In {\it1115}, the semi-major axis of the dark matter seems to be aligned with the semi-minor axis of the stellar component; it is the most strongly misaligned (anti-aligned) lens system of all lenses studied here. This may be explained by its external shear ($\gamma \sim 0.1$) emanating from its group environment (see Table \ref{tab:lensproperties}).

We performed tests for the robustness of our shape estimate and its dependence, in particular, on the shear prior. Only two systems, {\it1422} and {\it2045}, reacted to a stricter prior on the external shear $|\gamma|$. {\it1422} requires a rather large external shear ($|\gamma|\sim0.25$). Restricting the range of the allowed external shear values did not affect the misalignment of the dark matter and stellar mass distributions. It did, however, increase the ellipticity of the reconstructed mass distribution: we observed a trade-off between the galaxy's ellipticity and the magnitude of the external shear. {\it2045} on the other hand is particularly interesting. Allowing the shear to roam free as for the other lenses, favours an anomalously high shear ($\gamma > 0.5$), with strong misalignment. However, such models also produce spurious extra images along the arc. By limiting the maximum shear $|\gamma|$ to 0.1 the extra images are removed, though the lens then pushes on its maximum shear prior. We find that the mass map is robust against a stronger shear prior, and does not affect the alignment of the distributions. It produces well-aligned dark matter and stellar distributions. We believe that a strong shear prior for this particular lens is justified by the appearance of spurious extra images if higher shear is allowed. We furthermore explore potential degeneracies between the estimation of the shape and the required external shear in the Appendix. As shown in Figure~\ref{fig:thetascatter} the degeneracy can be largely broken for most lenses.

Figure~\ref{fig:wedgesradii} is similar to Figure~\ref{fig:wedgesall} with the difference that shapes are now averaged within different multiples of $R_e$. As the inertia tensor defined in Eq.~\ref{eq:inertiatensor} scales depends on the square of the radius, averaging up to different limiting radii could potentially yield different results. We observe that the shape estimator is robust to these changes. However, we note that there appears to be a slight trend that the ellipticity of the stellar distribution increases with increasing radius of enclosure. Whether this trend persists with data of a higher quality remains to be seen. There is potentially a similar trend for the dark matter distribution, although the samples are consistent with a constant ellipticity as well. {\it0957}, and to some degree {\it1422}, display, however, some signs of isophotal twist complicating the interpretation (see Appendix \ref{sec:shearshapedeg}).

\section{Discussion}\label{sec:discussion}

\subsection{Consistency with weak lensing estimates}
In contrast to strong lensing, weak lensing as a probe cannot constrain $e_{dm}$ directly. Rather, it is sensitive to the ratio $f_{h} = (e_{dm}/e_{*})\cdot\mathrm{cos} 2\Delta\theta$, where $\Delta\theta$ is the misalignment angle between the two distributions, and $e_{dm}\cdot\mathrm{cos} 2\Delta\theta$ the component of the dark matter ellipticity projected along the light distribution. In case of perfect alignment and identical shape, $f_{h}$ is equal to 1. In practice however, $f_{h}$ deviates from 1, as among other effects misalignment of the two distributions and different ellipticities, which cannot be disentangled, change its value.

We compare our results presented in the previous Section with the most recent study of the shape and alignment of the dark matter halo relative to the distribution of luminous matter by \citet{2015arXiv150704301S}. They perform a measurement on CFHTLenS data and on data from the Millennium Simulation \citep{2005Natur.435..629S} with a ray-tracing through by \citet{2009A&A...499...31H}. The constraints on the early-type galaxies in the highest mass bin ($\mathrm{log_{10}}M_{*}>11$) on CFHTLenS data are $f_{h}=-0.04^{+0.25}_{-0.25}$. On the simulated data, including cosmic shear and misalignments between the luminous and the dark matter, the constraints in the highest mass bin and for redshifts $0.4<z<0.6$ are $f_{h} = 0.359^{+0.011}_{-0.010}$. Due to a narrower intrinsic ellipticity distribution in the simulated data however, the latter constraints need to be corrected for with a factor $\sim$1/1.46. Our results are broadly consistent with both constraints, with the exception of the galaxy merger system \textit{1608} and the interesting and highly misaligned lens \textit{1115}. Our sample is, however, not large enough for a quantitative comparison.

\subsection{The lack of isolated misaligned lenses}
Our lens sample is small -- just 11 strong lensing systems -- yet it is interesting that we find no lens with strong misalignment and weak external shear ($\gamma < 0.1$). This is perhaps to be expected. \citet{1979ApJ...233..872H} were the first to study the stability of orbits within triaxial figures, likely the physical situation in real elliptical galaxies. They found that while stable periodic orbits exist in all three symmetry planes of a triaxial figure, tube orbits about the intermediate axis are unstable. \citet{1985MNRAS.215..731D} studied the case of orbits misaligned to all three symmetry planes. They found that while misaligned stable orbits can be found, they exit only for limited ranges of the Hamiltonian. \citet{1988A&A...206..269M} extended these results to rotating triaxial figures showing that once rotation is included such orbits become even rarer. These early results are supported by more recent numerical work. \citet{2013MNRAS.434.2971D} show that discs can never form perpendicular to the intermediate axis, in good agreement with earlier work by \citet{1979ApJ...233..872H}. However, they find that discs can survive off-axis if gas is present, since gas cooling can maintain the misalignment. \citet{2015MNRAS.452.4094D} go on to show that if this gas supply is switched off (as is the case for giant elliptical galaxies), then continuing collisionless minor mergers/interactions bring the disc rapidly into alignment with its host halo. Taken together, these results suggest that alignment between light and dark should be the norm for isolated gas-free triaxial galaxies, where here we define `isolated' as having external shear $\gamma < 0.1$.

\section{Conclusions}\label{sec:conclusions}
We have measured the projected radial density profile, shape and alignment of the stellar and dark matter distributions in 11 lens systems using a new non-parametric lens tool, \Glass. We focussed on lenses that have either time delay data or stellar mass maps that contribute significantly to the central potential. In a previous paper, we showed that data of this quality are required to determine the projected shape of dark matter haloes \citep{2014MNRAS.445.2181C}. We measured the shape and alignment using the eigenvalues $\lambda_i$ and eigenvectors of the 2D moment of inertia tensor of the stellar and dark matter distributions, defining an estimator of the ellipticity $e = (\lambda_{1}-\lambda_{2})/(\lambda_{1}-\lambda_{2})$ ($\lambda_{1} \geq \lambda_{2}$; see Section~\ref{sec:shapemethod}). We averaged $e$ over the range 1-5$R_e$, where $R_e$ is the effective radius of the light profile.

Our key results are as follows:

\begin{itemize}
\item The projected dark matter density profile in these systems -- under the assumption of a Chabrier stellar initial mass function -- shows significant variation from galaxy to galaxy. Those with an outermost image beyond $\sim 10$\,kpc are very well fit by a projected NFW profile; those with images within 10\,kpc appear to be more concentrated than NFW, as expected if their dark haloes contract due to baryonic cooling. There could be several explanations for this behaviour. It could be a selection effect where lower mass galaxies (expected to have closer-in images) can only strong lens if they are substantially contracted; it could owe to more massive haloes (with further out images) being less affected by baryons; or it could be that we simply cannot yet resolve the contraction at the centre of the more massive systems. We will explore such questions in a forthcoming paper.

\item The dark matter haloes of all lenses in our sample are rounder than the light distribution over the range $1R_e < R < 5R_e$. As we average over larger radii, there is a slight trend for the lenses, except the ones with a large stellar ellipticity, to become increasingly elliptical in their stellar distributions. A similar behaviour can be seen also for the dark matter, though the effect is not statistically significant. The dark matter haloes are never more elliptical than $e_{dm} = 0.2$, while their stars can extend to $e_* > 0.2$.

\item Three systems have a high stellar ellipticity ($e_* > 0.25$) and correspondingly high alignment between light and dark. One of these -- {\it1608} -- is a known merging pair. We speculate that the other two lens systems ({\it0712} and {\it2016}) may also be recent post-merger systems. 

\item Galaxies with high dark matter ellipticity and weak external shear ($\gamma < 0.1$) show strong alignment; those with strong shear ($\gamma \gtrsim 0.1$) can be highly misaligned. This is reassuring since isolated misaligned galaxies are expected to be unstable \citep[e.g.][]{1979ApJ...233..872H,1988A&A...206..269M,2007ApJ...670.1027A,2015arXiv150203429D}. We find that lenses with external shear $\gamma < 0.1$ appear to be sufficiently isolated that their luminous and dark matter distributions are well-aligned.

\end{itemize}

Our results on the shape and radial profile of the dark matter haloes in these lenses provide new constraints on galaxy formation models. Dark matter haloes can contract due to baryonic cooling \citep[e.g.][]{1986ApJ...301...27B} or expand due to energetic feedback from supernovae \citep[e.g.][]{2005MNRAS.356..107R,2012MNRAS.421.3464P,2014Natur.506..171P,2015arXiv150202036O,2015arXiv150804143R} or AGN \citep[e.g.][]{2012MNRAS.422.3081M}. The latest simulations for elliptical galaxies forming in $\Lambda$CDM without AGN feedback suggest that contraction should win \citep{2015MNRAS.453.2447D}. Under the assumption of a Chabrier IMF, our results imply that AGN do not play a major role in the lower mass lenses in our sample (since these are contracted with respect to NFW), but perhaps do act to counterbalance contraction in the more massive lenses. For the shape and alignment of lens galaxies, the models must reproduce rather round dark matter halos within $5R_e$, and misaligned stellar and dark matter distributions in lenses with strong external shear ($\gamma > 0.1$). It will be interesting to see whether the latest galaxy formation models reproduce this naturally, or it poses a challenge. Finally, we note that strong misalignments may be problematic for those alternative gravity theories in which light is the only source of gravity, since in such models light and dark must necessarily be highly correlated.

\section{Acknowledgements}\label{sec:acknowledgements}
We thank Simon Birrer and William Hartley for useful discussion. We also thank the anonymous referee for constructive comments. JIR would like to acknowledge support from SNF grant PP00P2\_128540/1. DL's research is part of the project GLENCO, funded under the European Seventh Framework Programme, Ideas, Grant Agreement n. 259349.

\bibliographystyle{mn2e}
\bibliography{LightVsDarkSLGals}

\begin{thebibliography}{114}
\expandafter\ifx\csname natexlab\endcsname\relax\def\natexlab#1{#1}\fi

\bibitem[{{Adams} {et~al}\mbox{.}(2007){Adams}, {Bloch}, {Butler}, {Druce}, \&
  {Ketchum}}]{2007ApJ...670.1027A}
{Adams} F.~C., {Bloch} A.~M., {Butler} S.~C., {Druce} J.~M., {Ketchum} J.~A.,
  2007, \apj, 670, 1027

\bibitem[{{Angonin-Willaime} {et~al}\mbox{.}(1994){Angonin-Willaime},
  {Soucail}, \& {Vanderriest}}]{1994A&A...291..411A}
{Angonin-Willaime} M.-C., {Soucail} G., {Vanderriest} C., 1994, \aap, 291, 411

\bibitem[{{Auger} {et~al}\mbox{.}(2007){Auger}, {Fassnacht}, {Abrahamse},
  {Lubin}, \& {Squires}}]{2007AJ....134..668A}
{Auger} M.~W., {Fassnacht} C.~D., {Abrahamse} A.~L., {Lubin} L.~M., {Squires}
  G.~K., 2007, \aj, 134, 668

\bibitem[{{Auger} {et~al}\mbox{.}(2010){Auger}, {Treu}, {Bolton}, {Gavazzi},
  {Koopmans}, {Marshall}, {Moustakas}, \& {Burles}}]{2010ApJ...724..511A}
{Auger} M.~W., {Treu} T., {Bolton} A.~S., {Gavazzi} R., {Koopmans} L.~V.~E.,
  {Marshall} P.~J., {Moustakas} L.~A., {Burles} S., 2010, \apj, 724, 511

\bibitem[{{Bade} {et~al}\mbox{.}(1997){Bade}, {Siebert}, {Lopez}, {Voges}, \&
  {Reimers}}]{1997A&A...317L..13B}
{Bade} N., {Siebert} J., {Lopez} S., {Voges} W., {Reimers} D., 1997, \aap, 317,
  L13

\bibitem[{{Baltz} {et~al}\mbox{.}(2009){Baltz}, {Marshall}, \&
  {Oguri}}]{2009JCAP...01..015B}
{Baltz} E.~A., {Marshall} P., {Oguri} M., 2009, \jcap, 1, 15

\bibitem[{{Barkana}(1997)}]{1997ApJ...489...21B}
{Barkana} R., 1997, \apj, 489, 21

\bibitem[{{Binney}(1978)}]{1978ComAp...8...27B}
{Binney} J., 1978, Comments on Astrophysics, 8, 27

\bibitem[{{Blandford} \& {Narayan}(1986)}]{1986ApJ...310..568B}
{Blandford} R., {Narayan} R., 1986, \apj, 310, 568

\bibitem[{{Blumenthal} {et~al}\mbox{.}(1986){Blumenthal}, {Faber}, {Flores}, \&
  {Primack}}]{1986ApJ...301...27B}
{Blumenthal} G.~R., {Faber} S.~M., {Flores} R., {Primack} J.~R., 1986, \apj,
  301, 27

\bibitem[{{Brainerd} \& {Wright}(2000)}]{2000astro.ph..6281B}
{Brainerd} T.~G., {Wright} C.~O., 2000, ArXiv Astrophysics e-prints

\bibitem[{{Bruzual} \& {Charlot}(2003)}]{2003MNRAS.344.1000B}
{Bruzual} G., {Charlot} S., 2003, \mnras, 344, 1000

\bibitem[{{Cappellari} {et~al}\mbox{.}(2013){Cappellari}, {McDermid},
  {Alatalo}, {Blitz}, {Bois}, {Bournaud}, {Bureau}, {Crocker}, {Davies},
  {Davis}, {de Zeeuw}, {Duc}, {Emsellem}, {Khochfar}, {Krajnovi{\'c}},
  {Kuntschner}, {Morganti}, {Naab}, {Oosterloo}, {Sarzi}, {Scott}, {Serra},
  {Weijmans}, \& {Young}}]{2013MNRAS.432.1862C}
{Cappellari} M. {et~al.}, 2013, \mnras, 432, 1862

\bibitem[{{Chabrier}(2003)}]{2003PASP..115..763C}
{Chabrier} G., 2003, \pasp, 115, 763

\bibitem[{{Chartas} {et~al}\mbox{.}(1998){Chartas}, {Chuss}, {Forman}, {Jones},
  \& {Shapiro}}]{1998ApJ...504..661C}
{Chartas} G., {Chuss} D., {Forman} W., {Jones} C., {Shapiro} I., 1998, \apj,
  504, 661

\bibitem[{{Chartas} {et~al}\mbox{.}(2002){Chartas}, {Gupta}, {Garmire},
  {Jones}, {Falco}, {Shapiro}, \& {Tavecchio}}]{2002ApJ...565...96C}
{Chartas} G., {Gupta} V., {Garmire} G., {Jones} C., {Falco} E.~E., {Shapiro}
  I.~I., {Tavecchio} F., 2002, \apj, 565, 96

\bibitem[{{Clampitt} \& {Jain}(2015)}]{2015arXiv150603536C}
{Clampitt} J., {Jain} B., 2015, ArXiv e-prints

\bibitem[{{Coles}(2008)}]{2008ApJ...679...17C}
{Coles} J., 2008, \apj, 679, 17

\bibitem[{{Coles} {et~al}\mbox{.}(2014){Coles}, {Read}, \&
  {Saha}}]{2014MNRAS.445.2181C}
{Coles} J.~P., {Read} J.~I., {Saha} P., 2014, \mnras, 445, 2181

\bibitem[{{Curran} {et~al}\mbox{.}(2011){Curran}, {Whiting}, {Tanna},
  {Bignell}, \& {Webb}}]{2011MNRAS.413L..86C}
{Curran} S.~J., {Whiting} M.~T., {Tanna} A., {Bignell} C., {Webb} J.~K., 2011,
  \mnras, 413, L86

\bibitem[{{Dai} \& {Kochanek}(2005)}]{2005ApJ...625..633D}
{Dai} X., {Kochanek} C.~S., 2005, \apj, 625, 633

\bibitem[{{de Zeeuw}(1985)}]{1985MNRAS.215..731D}
{de Zeeuw} T., 1985, \mnras, 215, 731

\bibitem[{{Debattista} {et~al}\mbox{.}(2008){Debattista}, {Moore}, {Quinn},
  {Kazantzidis}, {Maas}, {Mayer}, {Read}, \& {Stadel}}]{2007arXiv0707.0737D}
{Debattista} V.~P., {Moore} B., {Quinn} T., {Kazantzidis} S., {Maas} R.,
  {Mayer} L., {Read} J., {Stadel} J., 2008, \apj, 681, 1076

\bibitem[{{Debattista} {et~al}\mbox{.}(2013){Debattista}, {Ro{\v s}kar},
  {Valluri}, {Quinn}, {Moore}, \& {Wadsley}}]{2013MNRAS.434.2971D}
{Debattista} V.~P., {Ro{\v s}kar} R., {Valluri} M., {Quinn} T., {Moore} B.,
  {Wadsley} J., 2013, \mnras, 434, 2971

\bibitem[{{Debattista} {et~al}\mbox{.}(2015{\natexlab{a}}){Debattista}, {van
  den Bosch}, {Roskar}, {Quinn}, {Moore}, \& {Cole}}]{2015arXiv150203429D}
{Debattista} V.~P., {van den Bosch} F.~C., {Roskar} R., {Quinn} T., {Moore} B.,
  {Cole} D.~R., 2015{\natexlab{a}}, ArXiv e-prints

\bibitem[{{Debattista} {et~al}\mbox{.}(2015{\natexlab{b}}){Debattista}, {van
  den Bosch}, {Ro{\v s}kar}, {Quinn}, {Moore}, \& {Cole}}]{2015MNRAS.452.4094D}
{Debattista} V.~P., {van den Bosch} F.~C., {Ro{\v s}kar} R., {Quinn} T.,
  {Moore} B., {Cole} D.~R., 2015{\natexlab{b}}, \mnras, 452, 4094

\bibitem[{{Dominik}(1999)}]{1999A&A...349..108D}
{Dominik} M., 1999, \aap, 349, 108

\bibitem[{{Dubinski}(1994)}]{1994ApJ...431..617D}
{Dubinski} J., 1994, \apj, 431, 617

\bibitem[{{Dubinski} \& {Carlberg}(1991)}]{1991ApJ...378..496D}
{Dubinski} J., {Carlberg} R.~G., 1991, \apj, 378, 496

\bibitem[{{Dutton} {et~al}\mbox{.}(2015){Dutton}, {Macci{\`o}}, {Stinson},
  {Gutcke}, {Penzo}, \& {Buck}}]{2015MNRAS.453.2447D}
{Dutton} A.~A., {Macci{\`o}} A.~V., {Stinson} G.~S., {Gutcke} T.~A., {Penzo}
  C., {Buck} T., 2015, \mnras, 453, 2447

\bibitem[{{Fassnacht} {et~al}\mbox{.}(1999){Fassnacht}, {Blandford}, {Cohen},
  {Matthews}, {Pearson}, {Readhead}, {Womble}, {Myers}, {Browne}, {Jackson},
  {Marlow}, {Wilkinson}, {Koopmans}, {de Bruyn}, {Schilizzi}, {Bremer}, \&
  {Miley}}]{1999AJ....117..658F}
{Fassnacht} C.~D. {et~al.}, 1999, \aj, 117, 658

\bibitem[{{Fassnacht} \& {Cohen}(1998)}]{1998AJ....115..377F}
{Fassnacht} C.~D., {Cohen} J.~G., 1998, \aj, 115, 377

\bibitem[{{Fassnacht} {et~al}\mbox{.}(2006){Fassnacht}, {Gal}, {Lubin},
  {McKean}, {Squires}, \& {Readhead}}]{2006ApJ...642...30F}
{Fassnacht} C.~D., {Gal} R.~R., {Lubin} L.~M., {McKean} J.~P., {Squires} G.~K.,
  {Readhead} A.~C.~S., 2006, \apj, 642, 30

\bibitem[{{Fassnacht} \& {Lubin}(2002)}]{2002AJ....123..627F}
{Fassnacht} C.~D., {Lubin} L.~M., 2002, \aj, 123, 627

\bibitem[{{Fassnacht} {et~al}\mbox{.}(2002){Fassnacht}, {Xanthopoulos},
  {Koopmans}, \& {Rusin}}]{2002ApJ...581..823F}
{Fassnacht} C.~D., {Xanthopoulos} E., {Koopmans} L.~V.~E., {Rusin} D., 2002,
  \apj, 581, 823

\bibitem[{{Ferreras} {et~al}\mbox{.}(2013){Ferreras}, {La Barbera}, {de la
  Rosa}, {Vazdekis}, {de Carvalho}, {Falc{\'o}n-Barroso}, \&
  {Ricciardelli}}]{2013MNRAS.429L..15F}
{Ferreras} I., {La Barbera} F., {de la Rosa} I.~G., {Vazdekis} A., {de
  Carvalho} R.~R., {Falc{\'o}n-Barroso} J., {Ricciardelli} E., 2013, \mnras,
  429, L15

\bibitem[{{Ferreras} {et~al}\mbox{.}(2012){Ferreras}, {Mavromatos},
  {Sakellariadou}, \& {Yusaf}}]{2012PhRvD..86h3507F}
{Ferreras} I., {Mavromatos} N.~E., {Sakellariadou} M., {Yusaf} M.~F., 2012,
  \prd, 86, 083507

\bibitem[{{Ferreras} {et~al}\mbox{.}(2008){Ferreras}, {Saha}, \&
  {Burles}}]{2008MNRAS.383..857F}
{Ferreras} I., {Saha} P., {Burles} S., 2008, \mnras, 383, 857

\bibitem[{{Ferreras} {et~al}\mbox{.}(2010){Ferreras}, {Saha}, {Leier},
  {Courbin}, \& {Falco}}]{2010MNRAS.409L..30F}
{Ferreras} I., {Saha} P., {Leier} D., {Courbin} F., {Falco} E.~E., 2010,
  \mnras, 409, L30

\bibitem[{{Ferreras} {et~al}\mbox{.}(2005){Ferreras}, {Saha}, \&
  {Williams}}]{2005ApJ...623L...5F}
{Ferreras} I., {Saha} P., {Williams} L.~L.~R., 2005, \apjl, 623, L5

\bibitem[{{Garrett} {et~al}\mbox{.}(1992){Garrett}, {Walsh}, \&
  {Carswell}}]{1992MNRAS.254P..27G}
{Garrett} M.~A., {Walsh} D., {Carswell} R.~F., 1992, \mnras, 254, 27P

\bibitem[{{Grant} {et~al}\mbox{.}(2004){Grant}, {Bautz}, {Chartas}, \&
  {Garmire}}]{2004ApJ...610..686G}
{Grant} C.~E., {Bautz} M.~W., {Chartas} G., {Garmire} G.~P., 2004, \apj, 610,
  686

\bibitem[{{Heiligman} \& {Schwarzschild}(1979)}]{1979ApJ...233..872H}
{Heiligman} G., {Schwarzschild} M., 1979, \apj, 233, 872

\bibitem[{{Helmi}(2004)}]{2004ApJ...610L..97H}
{Helmi} A., 2004, \apjl, 610, L97

\bibitem[{{Hilbert} {et~al}\mbox{.}(2009){Hilbert}, {Hartlap}, {White}, \&
  {Schneider}}]{2009A&A...499...31H}
{Hilbert} S., {Hartlap} J., {White} S.~D.~M., {Schneider} P., 2009, \aap, 499,
  31

\bibitem[{{Hirata} \& {Seljak}(2004)}]{2004PhRvD..70f3526H}
{Hirata} C.~M., {Seljak} U., 2004, \prd, 70, 063526

\bibitem[{{Hjorth} {et~al}\mbox{.}(2002){Hjorth}, {Burud}, {Jaunsen},
  {Schechter}, {Kneib}, {Andersen}, {Korhonen}, {Clasen}, {Kaas},
  {{\O}stensen}, {Pelt}, \& {Pijpers}}]{2002ApJ...572L..11H}
{Hjorth} J. {et~al.}, 2002, \apjl, 572, L11

\bibitem[{{Hoekstra} {et~al}\mbox{.}(2004){Hoekstra}, {Yee}, \&
  {Gladders}}]{2004ApJ...606...67H}
{Hoekstra} H., {Yee} H.~K.~C., {Gladders} M.~D., 2004, \apj, 606, 67

\bibitem[{{Ibata} {et~al}\mbox{.}(2001){Ibata}, {Lewis}, {Irwin}, {Totten}, \&
  {Quinn}}]{2001ApJ...551..294I}
{Ibata} R., {Lewis} G.~F., {Irwin} M., {Totten} E., {Quinn} T., 2001, \apj,
  551, 294

\bibitem[{{Jackson} {et~al}\mbox{.}(1998){Jackson}, {Nair}, {Browne},
  {Wilkinson}, {Muxlow}, {de Bruyn}, {Koopmans}, {Bremer}, {Snellen}, {Miley},
  {Schilizzi}, {Myers}, {Fassnacht}, {Womble}, {Readhead}, {Blandford}, \&
  {Pearson}}]{1998MNRAS.296..483J}
{Jackson} N. {et~al.}, 1998, \mnras, 296, 483

\bibitem[{{Jing} \& {Suto}(2002)}]{2002ApJ...574..538J}
{Jing} Y.~P., {Suto} Y., 2002, \apj, 574, 538

\bibitem[{{Katz} \& {Gunn}(1991)}]{1991ApJ...377..365K}
{Katz} N., {Gunn} J.~E., 1991, \apj, 377, 365

\bibitem[{{Kazantzidis} {et~al}\mbox{.}(2004){Kazantzidis}, {Kravtsov},
  {Zentner}, {Allgood}, {Nagai}, \& {Moore}}]{2004ApJ...611L..73K}
{Kazantzidis} S., {Kravtsov} A.~V., {Zentner} A.~R., {Allgood} B., {Nagai} D.,
  {Moore} B., 2004, \apjl, 611, L73

\bibitem[{{Keeton}(2001)}]{2001astro.ph..2341K}
{Keeton} C.~R., 2001, ArXiv: astro-ph/0102341

\bibitem[{{Keeton} {et~al}\mbox{.}(1998){Keeton}, {Kochanek}, \&
  {Falco}}]{1998ApJ...509..561K}
{Keeton} C.~R., {Kochanek} C.~S., {Falco} E.~E., 1998, \apj, 509, 561

\bibitem[{{Klypin} {et~al}\mbox{.}(2011){Klypin}, {Trujillo-Gomez}, \&
  {Primack}}]{2011ApJ...740..102K}
{Klypin} A.~A., {Trujillo-Gomez} S., {Primack} J., 2011, \apj, 740, 102

\bibitem[{{Kneib} {et~al}\mbox{.}(2000){Kneib}, {Cohen}, \&
  {Hjorth}}]{2000ApJ...544L..35K}
{Kneib} J.-P., {Cohen} J.~G., {Hjorth} J., 2000, \apjl, 544, L35

\bibitem[{{Kochanek} {et~al}\mbox{.}(2000){Kochanek}, {Falco}, {Impey},
  {Leh{\'a}r}, {McLeod}, {Rix}, {Keeton}, {Mu{\~n}oz}, \&
  {Peng}}]{2000ApJ...543..131K}
{Kochanek} C.~S. {et~al.}, 2000, \apj, 543, 131

\bibitem[{{Koopmans} {et~al}\mbox{.}(2006){Koopmans}, {Treu}, {Bolton},
  {Burles}, \& {Moustakas}}]{2006ApJ...649..599K}
{Koopmans} L.~V.~E., {Treu} T., {Bolton} A.~S., {Burles} S., {Moustakas} L.~A.,
  2006, \apj, 649, 599

\bibitem[{{K{\"u}ng} {et~al}\mbox{.}(2015){K{\"u}ng}, {Saha}, {More}, {Baeten},
  {Coles}, {Cornen}, {Macmillan}, {Marshall}, {More}, {Odermatt}, {Verma}, \&
  {Wilcox}}]{2015MNRAS.447.2170K}
{K{\"u}ng} R. {et~al.}, 2015, \mnras, 447, 2170

\bibitem[{{Lawrence} {et~al}\mbox{.}(1984){Lawrence}, {Schneider}, {Schmidt},
  {Bennett}, {Hewitt}, {Burke}, {Turner}, \& {Gunn}}]{1984Sci...223...46L}
{Lawrence} C.~R., {Schneider} D.~P., {Schmidt} M., {Bennett} C.~L., {Hewitt}
  J.~N., {Burke} B.~F., {Turner} E.~L., {Gunn} J.~E., 1984, Science, 223, 46

\bibitem[{{Leier} {et~al}\mbox{.}(2012){Leier}, {Ferreras}, \&
  {Saha}}]{2012MNRAS.424..104L}
{Leier} D., {Ferreras} I., {Saha} P., 2012, \mnras, 424, 104

\bibitem[{{Leier} {et~al}\mbox{.}(2011){Leier}, {Ferreras}, {Saha}, \&
  {Falco}}]{2011ApJ...740...97L}
{Leier} D., {Ferreras} I., {Saha} P., {Falco} E.~E., 2011, \apj, 740, 97

\bibitem[{{Lubini} \& {Coles}(2012)}]{2012MNRAS.425.3077L}
{Lubini} M., {Coles} J., 2012, \mnras, 425, 3077

\bibitem[{{Lux} {et~al}\mbox{.}(2012){Lux}, {Read}, {Lake}, \&
  {Johnston}}]{2012MNRAS.424L..16L}
{Lux} H., {Read} J.~I., {Lake} G., {Johnston} K.~V., 2012, \mnras, 424, L16

\bibitem[{{Macci{\`o}} {et~al}\mbox{.}(2007){Macci{\`o}}, {Dutton}, {van den
  Bosch}, {Moore}, {Potter}, \& {Stadel}}]{2007MNRAS.378...55M}
{Macci{\`o}} A.~V., {Dutton} A.~A., {van den Bosch} F.~C., {Moore} B., {Potter}
  D., {Stadel} J., 2007, \mnras, 378, 55

\bibitem[{{Mandelbaum} {et~al}\mbox{.}(2006){Mandelbaum}, {Hirata},
  {Broderick}, {Seljak}, \& {Brinkmann}}]{2006MNRAS.370.1008M}
{Mandelbaum} R., {Hirata} C.~M., {Broderick} T., {Seljak} U., {Brinkmann} J.,
  2006, \mnras, 370, 1008

\bibitem[{{Mart{\'{\i}}n-Navarro} {et~al}\mbox{.}(2015){Mart{\'{\i}}n-Navarro},
  {Barbera}, {Vazdekis}, {Falc{\'o}n-Barroso}, \&
  {Ferreras}}]{2015MNRAS.447.1033M}
{Mart{\'{\i}}n-Navarro} I., {Barbera} F.~L., {Vazdekis} A.,
  {Falc{\'o}n-Barroso} J., {Ferreras} I., 2015, \mnras, 447, 1033

\bibitem[{{Martinet} \& {de Zeeuw}(1988)}]{1988A&A...206..269M}
{Martinet} L., {de Zeeuw} T., 1988, \aap, 206, 269

\bibitem[{{Martizzi} {et~al}\mbox{.}(2012){Martizzi}, {Teyssier}, {Moore}, \&
  {Wentz}}]{2012MNRAS.422.3081M}
{Martizzi} D., {Teyssier} R., {Moore} B., {Wentz} T., 2012, \mnras, 422, 3081

\bibitem[{{McKean} {et~al}\mbox{.}(2007){McKean}, {Koopmans}, {Flack},
  {Fassnacht}, {Thompson}, {Matthews}, {Blandford}, {Readhead}, \&
  {Soifer}}]{2007MNRAS.378..109M}
{McKean} J.~P. {et~al.}, 2007, \mnras, 378, 109

\bibitem[{{Momcheva} {et~al}\mbox{.}(2006){Momcheva}, {Williams}, {Keeton}, \&
  {Zabludoff}}]{2006ApJ...641..169M}
{Momcheva} I., {Williams} K., {Keeton} C., {Zabludoff} A., 2006, \apj, 641, 169

\bibitem[{{More} {et~al}\mbox{.}(2009){More}, {McKean}, {More}, {Porcas},
  {Koopmans}, \& {Garrett}}]{2009MNRAS.394..174M}
{More} A., {McKean} J.~P., {More} S., {Porcas} R.~W., {Koopmans} L.~V.~E.,
  {Garrett} M.~A., 2009, \mnras, 394, 174

\bibitem[{{Morgan} {et~al}\mbox{.}(2001){Morgan}, {Chartas}, {Malm}, {Bautz},
  {Burud}, {Hjorth}, {Jones}, \& {Schechter}}]{2001ApJ...555....1M}
{Morgan} N.~D., {Chartas} G., {Malm} M., {Bautz} M.~W., {Burud} I., {Hjorth}
  J., {Jones} S.~E., {Schechter} P.~L., 2001, \apj, 555, 1

\bibitem[{{Mortlock} \& {Turner}(2001)}]{2001MNRAS.327..552M}
{Mortlock} D.~J., {Turner} E.~L., 2001, \mnras, 327, 552

\bibitem[{{Myers} {et~al}\mbox{.}(1995){Myers}, {Fassnacht}, {Djorgovski},
  {Blandford}, {Matthews}, {Neugebauer}, {Pearson}, {Readhead}, {Smith},
  {Thompson}, {Womble}, {Browne}, {Wilkinson}, {Nair}, {Jackson}, {Snellen},
  {Miley}, {de Bruyn}, \& {Schilizzi}}]{1995ApJ...447L...5M}
{Myers} S.~T. {et~al.}, 1995, \apjl, 447, L5

\bibitem[{{Natarajan} \& {Refregier}(2000)}]{2000ApJ...538L.113N}
{Natarajan} P., {Refregier} A., 2000, \apjl, 538, L113

\bibitem[{{Navarro} {et~al}\mbox{.}(1996){Navarro}, {Frenk}, \&
  {White}}]{1996ApJ...462..563N}
{Navarro} J.~F., {Frenk} C.~S., {White} S.~D.~M., 1996, \apj, 462, 563

\bibitem[{{O{\~n}orbe} {et~al}\mbox{.}(2015){O{\~n}orbe}, {Boylan-Kolchin},
  {Bullock}, {Hopkins}, {Ker{\v e}s}, {Faucher-Gigu{\`e}re}, {Quataert}, \&
  {Murray}}]{2015arXiv150202036O}
{O{\~n}orbe} J., {Boylan-Kolchin} M., {Bullock} J.~S., {Hopkins} P.~F., {Ker{\v
  e}s} D., {Faucher-Gigu{\`e}re} C.-A., {Quataert} E., {Murray} N., 2015, ArXiv
  e-prints

\bibitem[{{Parker} {et~al}\mbox{.}(2007){Parker}, {Hoekstra}, {Hudson}, {van
  Waerbeke}, \& {Mellier}}]{2007ApJ...669...21P}
{Parker} L.~C., {Hoekstra} H., {Hudson} M.~J., {van Waerbeke} L., {Mellier} Y.,
  2007, \apj, 669, 21

\bibitem[{{Patnaik} {et~al}\mbox{.}(1992){Patnaik}, {Browne}, {Walsh},
  {Chaffee}, \& {Foltz}}]{1992MNRAS.259P...1P}
{Patnaik} A.~R., {Browne} I.~W.~A., {Walsh} D., {Chaffee} F.~H., {Foltz} C.~B.,
  1992, \mnras, 259, 1P

\bibitem[{{Patnaik} \& {Narasimha}(2001)}]{2001MNRAS.326.1403P}
{Patnaik} A.~R., {Narasimha} D., 2001, \mnras, 326, 1403

\bibitem[{{Peng} {et~al}\mbox{.}(2002){Peng}, {Ho}, {Impey}, \&
  {Rix}}]{2002AJ....124..266P}
{Peng} C.~Y., {Ho} L.~C., {Impey} C.~D., {Rix} H.-W., 2002, \aj, 124, 266

\bibitem[{{Pontzen} \& {Governato}(2012)}]{2012MNRAS.421.3464P}
{Pontzen} A., {Governato} F., 2012, \mnras, 421, 3464

\bibitem[{{Pontzen} \& {Governato}(2014)}]{2014Natur.506..171P}
{Pontzen} A., {Governato} F., 2014, \nat, 506, 171

\bibitem[{{Raychaudhury} {et~al}\mbox{.}(2003){Raychaudhury}, {Saha}, \&
  {Williams}}]{2003AJ....126...29R}
{Raychaudhury} S., {Saha} P., {Williams} L.~L.~R., 2003, \aj, 126, 29

\bibitem[{{Read}(2014)}]{2014JPhG...41f3101R}
{Read} J.~I., 2014, Journal of Physics G Nuclear Physics, 41, 063101

\bibitem[{{Read} {et~al}\mbox{.}(2015){Read}, {Agertz}, \&
  {Collins}}]{2015arXiv150804143R}
{Read} J.~I., {Agertz} O., {Collins} M.~L.~M., 2015, ArXiv e-prints

\bibitem[{{Read} \& {Gilmore}(2005)}]{2005MNRAS.356..107R}
{Read} J.~I., {Gilmore} G., 2005, \mnras, 356, 107

\bibitem[{{Read} \& {Moore}(2005)}]{2005MNRAS.361..971R}
{Read} J.~I., {Moore} B., 2005, \mnras, 361, 971

\bibitem[{{Saha} \& {Williams}(2004)}]{2004AJ....127.2604S}
{Saha} P., {Williams} L.~L.~R., 2004, \aj, 127, 2604

\bibitem[{{Schechter} {et~al}\mbox{.}(1997){Schechter}, {Bailyn}, {Barr},
  {Barvainis}, {Becker}, {Bernstein}, {Blakeslee}, {Bus}, {Dressler}, {Falco},
  {Fesen}, {Fischer}, {Gebhardt}, {Harmer}, {Hewitt}, {Hjorth}, {Hurt},
  {Jaunsen}, {Mateo}, {Mehlert}, {Richstone}, {Sparke}, {Thorstensen}, {Tonry},
  {Wegner}, {Willmarth}, \& {Worthey}}]{1997ApJ...475L..85S}
{Schechter} P.~L. {et~al.}, 1997, \apjl, 475, L85

\bibitem[{{Schechter} \& {Moore}(1993)}]{1993AJ....105....1S}
{Schechter} P.~L., {Moore} C.~B., 1993, \aj, 105, 1

\bibitem[{{Schechter} {et~al}\mbox{.}(2014){Schechter}, {Pooley}, {Blackburne},
  \& {Wambsganss}}]{2014ApJ...793...96S}
{Schechter} P.~L., {Pooley} D., {Blackburne} J.~A., {Wambsganss} J., 2014,
  \apj, 793, 96

\bibitem[{{Schneider} {et~al}\mbox{.}(1986){Schneider}, {Gunn}, {Turner},
  {Lawrence}, {Hewitt}, {Schmidt}, \& {Burke}}]{1986AJ.....91..991S}
{Schneider} D.~P., {Gunn} J.~E., {Turner} E.~L., {Lawrence} C.~R., {Hewitt}
  J.~N., {Schmidt} M., {Burke} B.~F., 1986, \aj, 91, 991

\bibitem[{{Schneider} {et~al}\mbox{.}(1992){Schneider}, {Ehlers}, \&
  {Falco}}]{1992grle.book.....S}
{Schneider} P., {Ehlers} J., {Falco} E.~E., 1992, {Gravitational Lenses}

\bibitem[{{Schrabback} {et~al}\mbox{.}(2015){Schrabback}, {Hilbert},
  {Hoekstra}, {Simon}, {van Uitert}, {Erben}, {Heymans}, {Hildebrandt},
  {Kitching}, {Mellier}, {Miller}, {Van Waerbeke}, {Bett}, {Coupon}, {Fu},
  {Hudson}, {Joachimi}, {Kilbinger}, \& {Kuijken}}]{2015arXiv150704301S}
{Schrabback} T. {et~al.}, 2015, ArXiv e-prints

\bibitem[{{Shalyapin} {et~al}\mbox{.}(2012){Shalyapin}, {Goicoechea}, \&
  {Gil-Merino}}]{2012A&A...540A.132S}
{Shalyapin} V.~N., {Goicoechea} L.~J., {Gil-Merino} R., 2012, \aap, 540, A132

\bibitem[{{Sluse} {et~al}\mbox{.}(2012){Sluse}, {Chantry}, {Magain}, {Courbin},
  \& {Meylan}}]{2012A&A...538A..99S}
{Sluse} D., {Chantry} V., {Magain} P., {Courbin} F., {Meylan} G., 2012, \aap,
  538, A99

\bibitem[{{Springel} {et~al}\mbox{.}(2005){Springel}, {White}, {Jenkins},
  {Frenk}, {Yoshida}, {Gao}, {Navarro}, {Thacker}, {Croton}, {Helly},
  {Peacock}, {Cole}, {Thomas}, {Couchman}, {Evrard}, {Colberg}, \&
  {Pearce}}]{2005Natur.435..629S}
{Springel} V. {et~al.}, 2005, \nat, 435, 629

\bibitem[{{Surpi} \& {Blandford}(2003)}]{2003ApJ...584..100S}
{Surpi} G., {Blandford} R.~D., 2003, \apj, 584, 100

\bibitem[{{Toft} {et~al}\mbox{.}(2003){Toft}, {Soucail}, \&
  {Hjorth}}]{2003MNRAS.344..337T}
{Toft} S., {Soucail} G., {Hjorth} J., 2003, \mnras, 344, 337

\bibitem[{{Tonry} \& {Kochanek}(1999)}]{1999AJ....117.2034T}
{Tonry} J.~L., {Kochanek} C.~S., 1999, \aj, 117, 2034

\bibitem[{{Treu} {et~al}\mbox{.}(2009){Treu}, {Gavazzi}, {Gorecki}, {Marshall},
  {Koopmans}, {Bolton}, {Moustakas}, \& {Burles}}]{2009ApJ...690..670T}
{Treu} T., {Gavazzi} R., {Gorecki} A., {Marshall} P.~J., {Koopmans} L.~V.~E.,
  {Bolton} A.~S., {Moustakas} L.~A., {Burles} S., 2009, \apj, 690, 670

\bibitem[{{Treu} {et~al}\mbox{.}(2006){Treu}, {Koopmans}, {Bolton}, {Burles},
  \& {Moustakas}}]{2006ApJ...640..662T}
{Treu} T., {Koopmans} L.~V., {Bolton} A.~S., {Burles} S., {Moustakas} L.~A.,
  2006, \apj, 640, 662

\bibitem[{{Tsvetkova} {et~al}\mbox{.}(2010){Tsvetkova}, {Vakulik}, {Shulga},
  {Schild}, {Dudinov}, {Minakov}, {Nuritdinov}, {Artamonov}, {Kochetov},
  {Smirnov}, {Sergeyev}, {Konichek}, {Sinelnikov}, {Zheleznyak}, {Bruevich},
  {Gaisin}, {Akhunov}, \& {Burkhonov}}]{2010MNRAS.406.2764T}
{Tsvetkova} V.~S. {et~al.}, 2010, \mnras, 406, 2764

\bibitem[{{van Uitert} {et~al}\mbox{.}(2012){van Uitert}, {Hoekstra},
  {Schrabback}, {Gilbank}, {Gladders}, \& {Yee}}]{2012A&A...545A..71V}
{van Uitert} E., {Hoekstra} H., {Schrabback} T., {Gilbank} D.~G., {Gladders}
  M.~D., {Yee} H.~K.~C., 2012, \aap, 545, A71

\bibitem[{{Warren} {et~al}\mbox{.}(1992){Warren}, {Quinn}, {Salmon}, \&
  {Zurek}}]{1992ApJ...399..405W}
{Warren} M.~S., {Quinn} P.~J., {Salmon} J.~K., {Zurek} W.~H., 1992, \apj, 399,
  405

\bibitem[{{Warren} {et~al}\mbox{.}(1996){Warren}, {Hewett}, {Lewis}, {Moller},
  {Iovino}, \& {Shaver}}]{1996MNRAS.278..139W}
{Warren} S.~J., {Hewett} P.~C., {Lewis} G.~F., {Moller} P., {Iovino} A.,
  {Shaver} P.~A., 1996, \mnras, 278, 139

\bibitem[{{Weymann} {et~al}\mbox{.}(1980){Weymann}, {Latham}, {Roger}, {Angel},
  {Green}, {Liebert}, {Turnshek}, {Turnshek}, \& {Tyson}}]{1980Natur.285..641W}
{Weymann} R.~J. {et~al.}, 1980, \nat, 285, 641

\bibitem[{{Wong} {et~al}\mbox{.}(2011){Wong}, {Keeton}, {Williams}, {Momcheva},
  \& {Zabludoff}}]{2011ApJ...726...84W}
{Wong} K.~C., {Keeton} C.~R., {Williams} K.~A., {Momcheva} I.~G., {Zabludoff}
  A.~I., 2011, \apj, 726, 84

\bibitem[{{Yee}(1988)}]{1988AJ.....95.1331Y}
{Yee} H.~K.~C., 1988, \aj, 95, 1331

\bibitem[{{Yee} \& {Ellingson}(1994)}]{1994AJ....107...28Y}
{Yee} H.~K.~C., {Ellingson} E., 1994, \aj, 107, 28

\bibitem[{{Yoo} {et~al}\mbox{.}(2005){Yoo}, {Kochanek}, {Falco}, \&
  {McLeod}}]{2005ApJ...626...51Y}
{Yoo} J., {Kochanek} C.~S., {Falco} E.~E., {McLeod} B.~A., 2005, \apj, 626, 51

\end{thebibliography}

\appendix
\section{Reconstructed Lenses}\label{sec:reconstructions}
\begin{table}
  \begin{center}
    \begin{tabular}{l r r r r}
      Lens & A [''] & B [''] & C [''] & D [''] \\ \hline
      0047 & 1.270 & -0.630 & 0.520 & -0.730 \\
           & 0.105 & -0.995 & -1.045 & 0.705 \\
      0414 & -0.472 & -1.061 & -1.1947 & 0.885 \\
           & 1.277 & -0.661 & -0.255 & -0.361 \\
      0712 & -0.013 & 0.795 & 0.747 & -0.391 \\
           & -0.804 & -0.156 & -0.292 & 0.307 \\
      0911 & 2.226 & -0.968 & -0.709 & -0.696 \\
           & 0.278 & -0.105 & -0.507 & 0.439 \\
      0957$^{a}$ & 1.408 & 0.182 & 2.860 & -1.540 \\
           & 5.034 & -1.018 & 3.470 & -0.050 \\
      1115 & 0.355 & -0.909 & -1.093 & 0.717 \\
           & 1.322 & -0.714 & -0.260 & -0.627 \\
      1422 & 1.079 & 0.357 & 0.742 & -0.205 \\
           & -0.095 & 0.973 & 0.656 & -0.147 \\
      1608 & -1.300 & -0.560 & -1.310 & 0.570 \\
           & -0.800 & 1.160 & 0.700 & -0.080 \\
      2016 & -1.735 & 0.335 & 0.437 & 1.268 \\
           & 1.778 & -1.450 & -1.435 & 0.276 \\
      2045 & 1.121 & 1.409 & 1.255 & -0.507 \\
           & 0.824 & 0.035 & 0.576 & -0.183 \\
      2237 & 0.598 & -0.075 & 0.791 & -0.710 \\
           & 0.758 & -0.939 & -0.411 & 0.271 \\
    \end{tabular}
    \caption[width=\linewidth]{The image positions of all the systems are listed in order of arrival time (image A has the shortest arrival time, image D the longest). The positions are relative to the centre of the lens galaxy. The first row contains the RA coordinate, the second the DEC coordinate of the image. \newline $^{a}$ The source lensed by {\it0957} has a second component. Images A and B are a double associated with the main galaxy component, images C and D with the lensed substructure.}
    \label{tab:lenspositions}
  \end{center}
\end{table}

\begin{figure*}
  \centering
  \includegraphics[width=.83\linewidth]{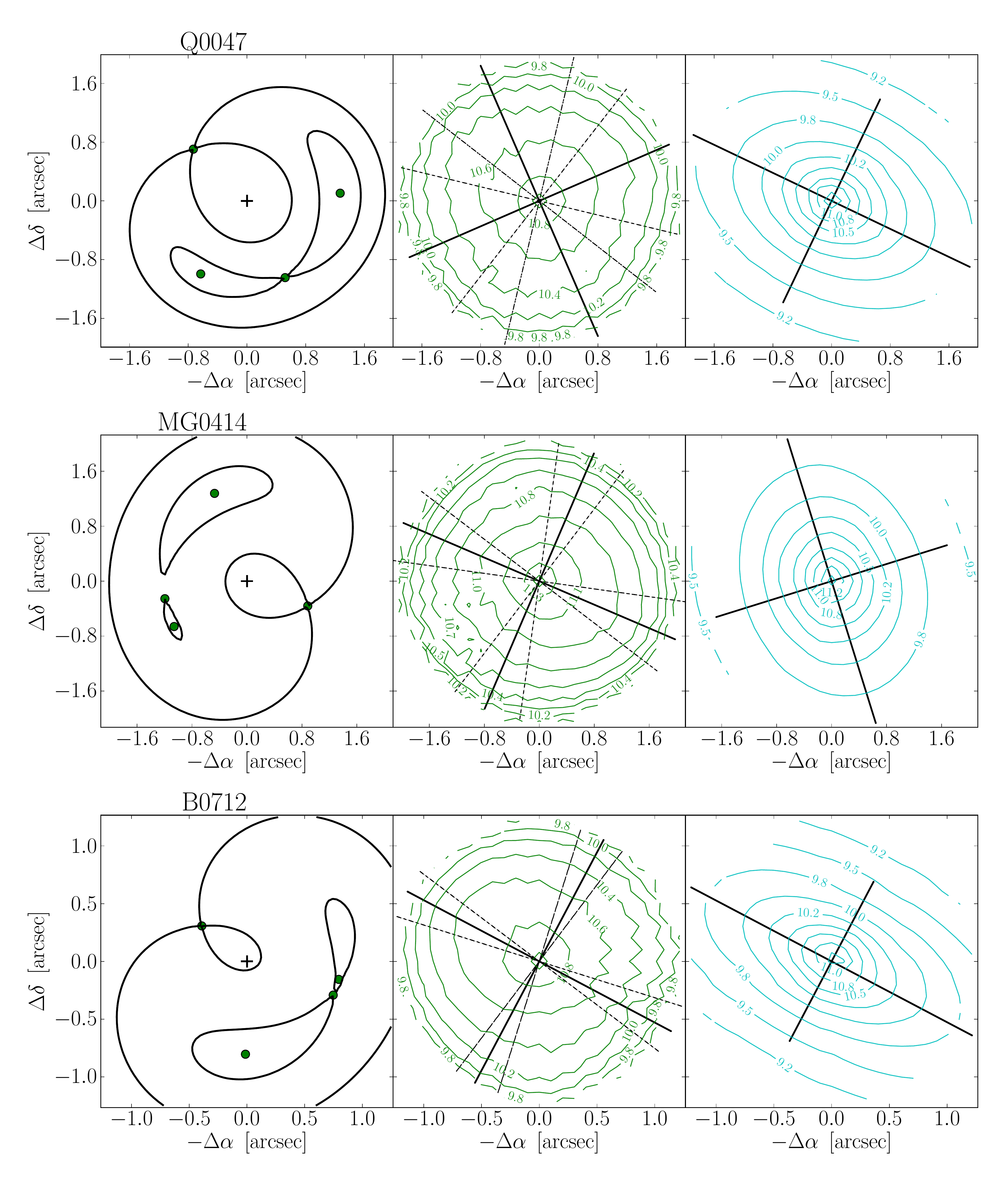}
  \caption[width=.65\linewidth]{The results of our reconstruction of each individual lens galaxy. The panels show, from left to right: the arrival time surface (the images are marked by the green circles); the surface mass density of the dark matter; and the surface mass density of the stars. The solid lines mark the eigenvalues and eigenvectors of the 2D moment of inertia tensor in each case; the dashed lines the 68\% confidence interval of these for the dark matter map.}
  \label{fig:lensreconstruction1}
\end{figure*}

\begin{figure*}
  \centering
  \includegraphics[width=.83\linewidth]{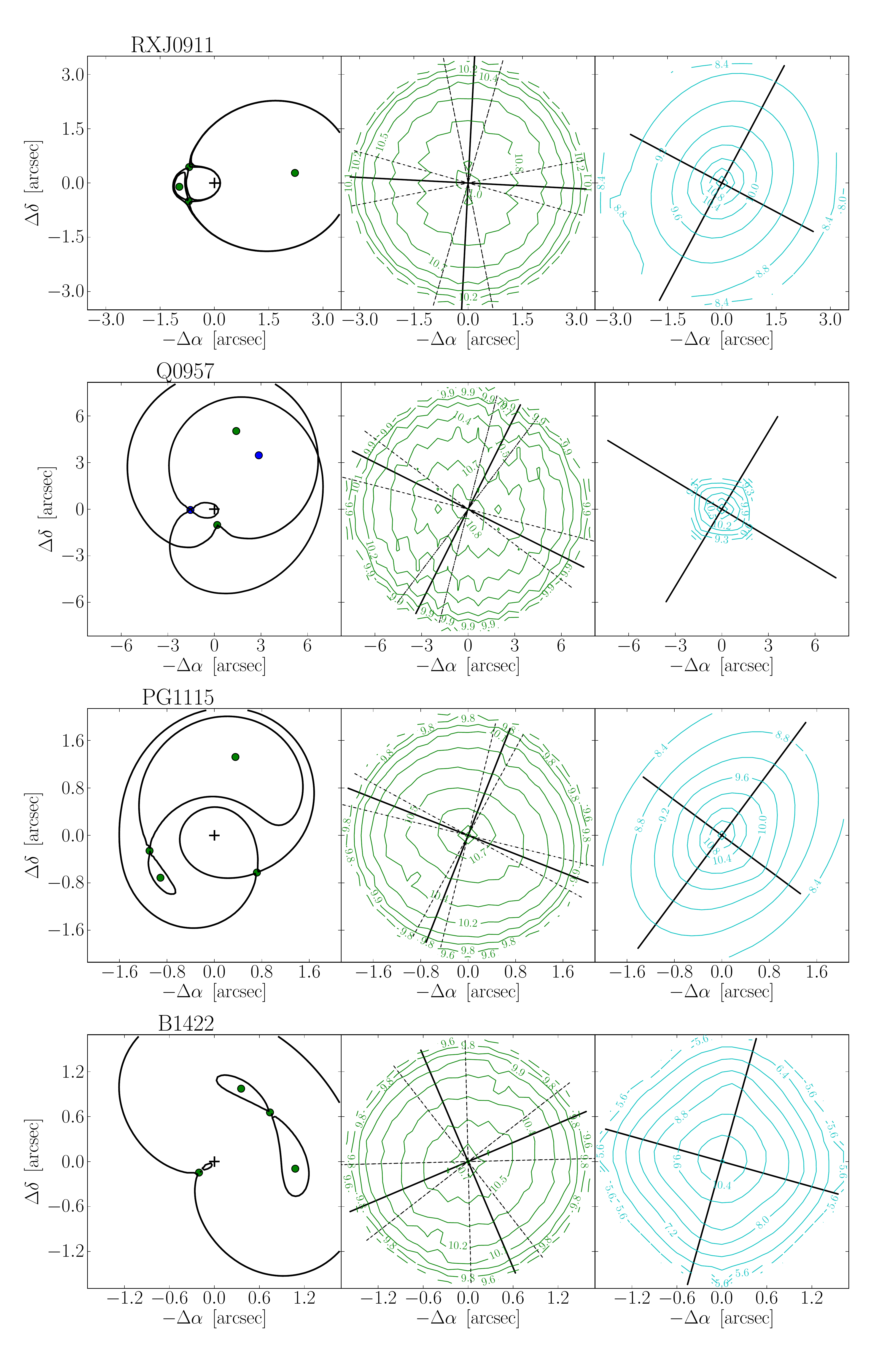}
  \caption[width=.65\linewidth]{The results of our reconstruction of each individual lens galaxy. Lines and symbols are as in Figure \ref{fig:lensreconstruction1}.}
  \label{fig:lensreconstruction2}
\end{figure*}

\begin{figure*}
  \centering
  \includegraphics[width=.83\linewidth]{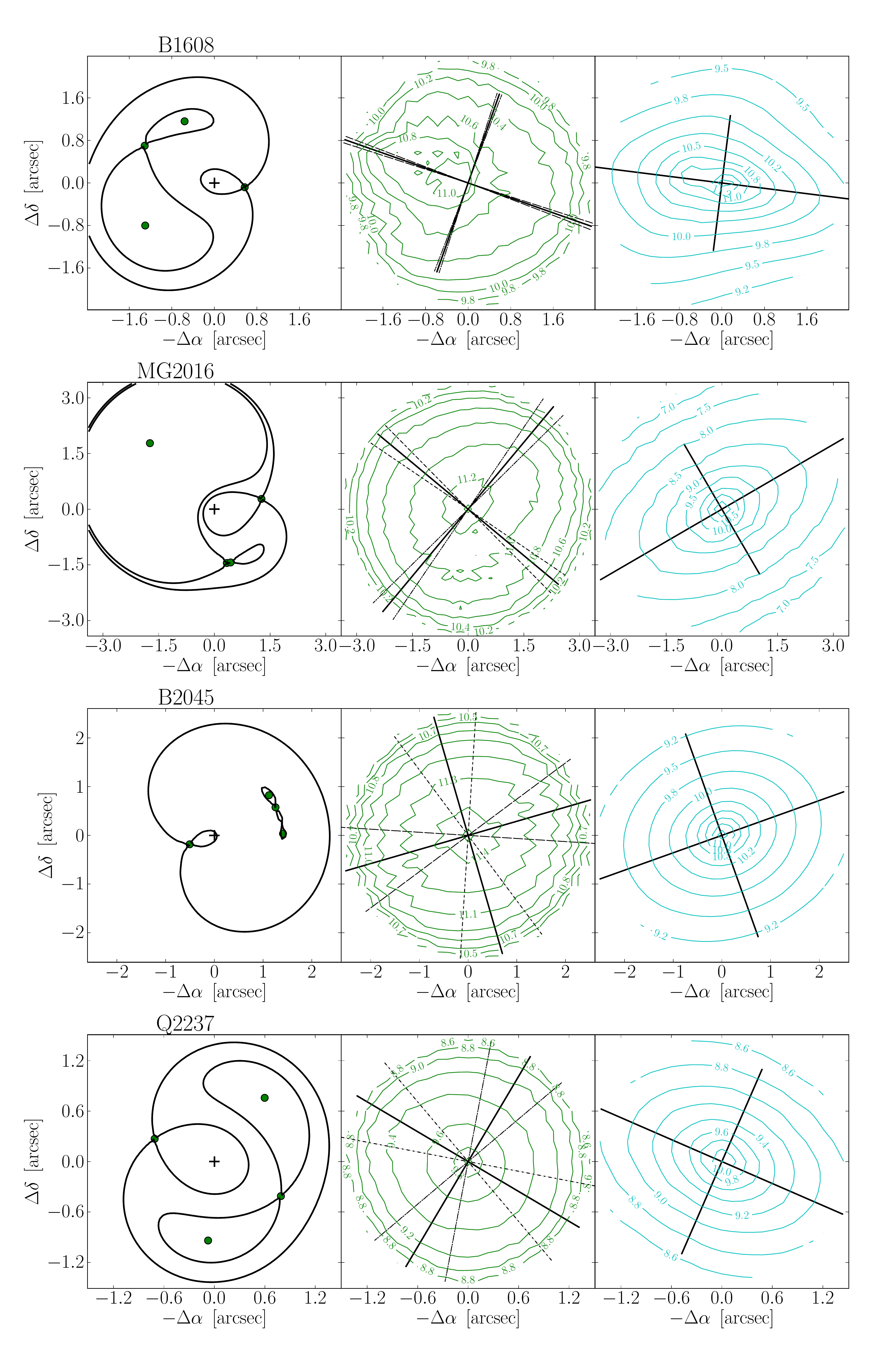}
  \caption[width=.65\linewidth]{The results of our reconstruction of each individual lens galaxy. Lines and symbols are as in Figure \ref{fig:lensreconstruction1}.}
  \label{fig:lensreconstruction3}
\end{figure*}

In this Appendix, we show the results of our lens modelling for each individual lens galaxy (Figures~\ref{fig:lensreconstruction1}-\ref{fig:lensreconstruction3}). The panels show, from left to right: the arrival time surface; the surface mass density of the dark matter; and the surface mass density of the stars. The solid lines mark the eigenvalues and eigenvectors of the 2D moment of inertia tensor in each case; the dashed lines denote the ranges 68\% of all the eigenvectors of the dark matter distribution corresponding to an individual model lie within. Figures~\ref{fig:wedgesall} \& \ref{fig:wedgesradii} are constructed from a combination of the eigenvalues (Eq.~\ref{eq:shapeestimate}), and the angle between the dark matter and stellar major axes (Eq.~\ref{eq:misalignment}). Note that the angular scale is always the same for the dark matter and stellar maps, but varies between the different lenses as marked on the Figure axes.

We note in Figure~\ref{fig:lensreconstruction2}, specifically for {\it0957} and to some degree also {\it1422}, twisting isophotes \citep[e.g.][]{1978ComAp...8...27B}.

\section{Degeneracy between $\theta_{dm}$ and $\theta_{g}$}\label{sec:shearshapedeg}
\begin{figure*}
  \centering
  \includegraphics[width=.93\linewidth]{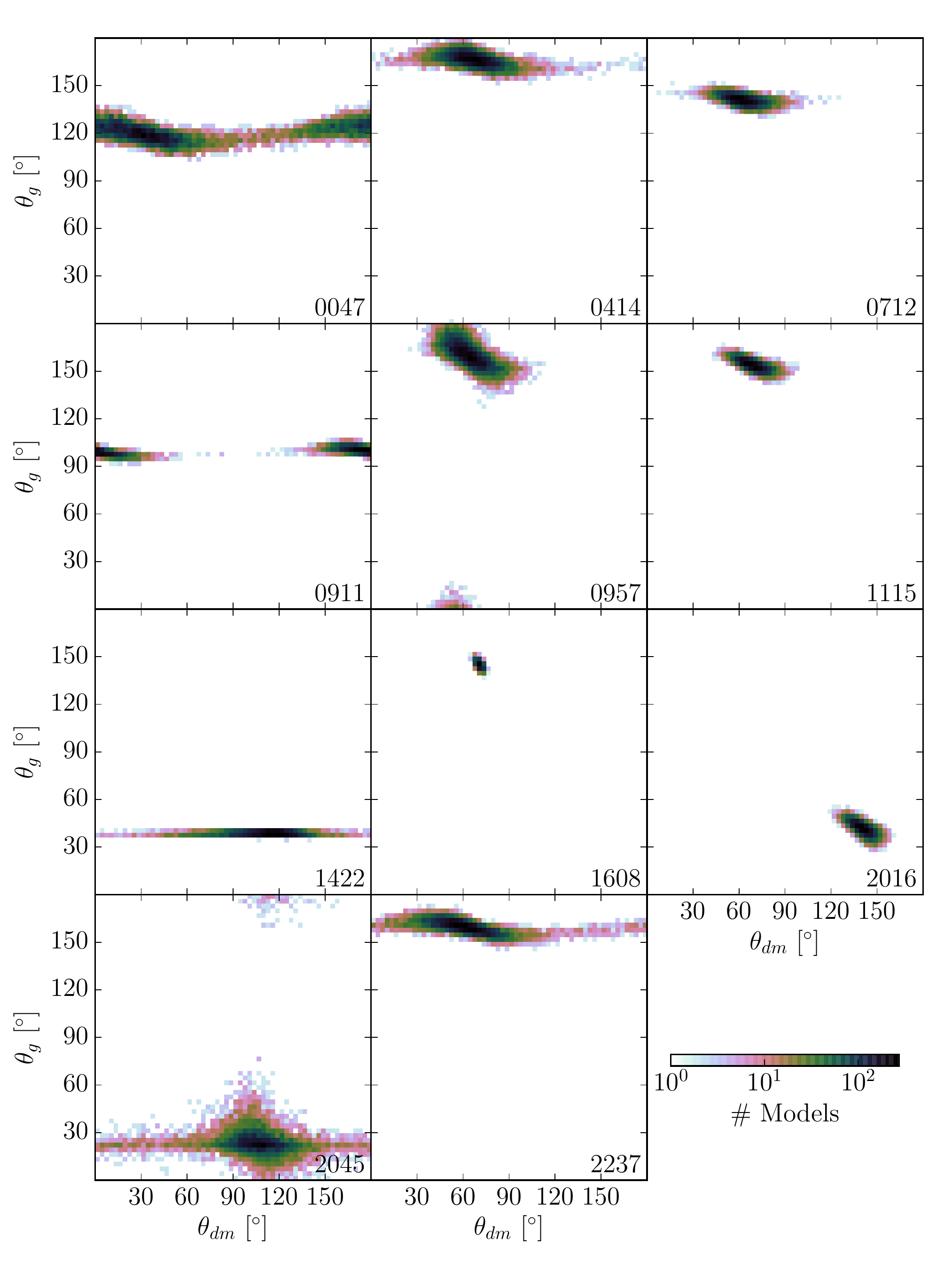}
  \caption[width=\linewidth]{The two-dimensional distribution of the position angles of the dark matter distribution $\theta_{dm}$ and the induced position angle of the required external shear $\theta_{g}$ of each reconstructed mass distribution. The panels display this distribution, which traces the degeneracy of the two parameters, for each reconstructed strong lens galaxy in the ensemble of solutions. Here, the angles are rotated to be measured north through east.}
  \label{fig:thetascatter}
\end{figure*}

Figure~\ref{fig:thetascatter} shows the degeneracy between the position angles of the dark matter halo $\theta_{dm}$ and the external shear $\theta_{g}$ for each strong lens galaxy. We note that the constraints on the direction of the external shear are stronger than on the dark matter distribution. There is a weak degeneracy between the two quantities, it can however for most lenses be largely broken. This is expected, since internal ellipticity and external shear tend to have similar effects, but are not an exact degeneracy \cite[cf.][]{1999A&A...349..108D}.

\end{document}